\documentclass{article}
\usepackage[utf8]{inputenc} % allow utf-8 input
\usepackage[T1]{fontenc}    % use 8-bit T1 fonts
\usepackage[preprint]{neurips_2024}
\usepackage{amsmath}
\usepackage[table,xcdraw]{xcolor}
\usepackage[most]{tcolorbox}
\usepackage{float}
\usepackage{amsfonts}       % blackboard math symbols
\usepackage{graphicx}
\usepackage{subcaption}
\usepackage{natbib}
\usepackage{amsthm}
\usepackage{setspace}
\usepackage{wrapfig}

\usepackage{hyperref}
\hypersetup{
    colorlinks=false,
    linkcolor=blue,
    filecolor=blue,      
    urlcolor=blue,
    pdftitle={Overleaf Example},
    pdfpagemode=FullScreen,
    }

\usepackage{booktabs}       % professional-quality tables
\usepackage{nicefrac}       % compact symbols for 1/2, etc.
\usepackage{microtype}      % microtypography
\usepackage{xcolor}         % colors
\usepackage[none]{hyphenat}

\usepackage{authblk}
\usepackage{multirow}

\usepackage[textsize=tiny]{todonotes}

\definecolor{chatglm2color}{HTML}{797979}
\definecolor{chatglm3color}{HTML}{4878D0}
\definecolor{llama2color}{HTML}{6ACC64}
\definecolor{baichuan2color}{HTML}{D5BB67}
\definecolor{claude1color}{HTML}{4878D0}
\definecolor{claude2color}{HTML}{956CB4}
\definecolor{palm2color}{HTML}{8C613C}
\definecolor{gpt3color}{HTML}{DC7EC0}
\definecolor{gpt4color}{HTML}{D65F5F}
\definecolor{enconarenacolor}{HTML}{C32136}

\newcommand{\chatglmtwo}{{\color{chatglm2color} \texttt{ChatGLM2}}}
\newcommand{\chatglmthree}{{\color{chatglm3color} \texttt{ChatGLM3}}}
\newcommand{\llama}{{\color{llama2color} \texttt{Llama2}}}
\newcommand{\baichuan}{{\color{baichuan2color} \texttt{Baichuan2}}}
\newcommand{\claudeone}{{\color{claude1color} \texttt{Claude1}}}
\newcommand{\claudetwo}{{\color{claude2color} \texttt{Claude2}}}
\newcommand{\palm}{{\color{palm2color} \texttt{PaLM}}}
\newcommand{\gptthree}{{\color{gpt3color} \texttt{GPT3.5}}}
\newcommand{\gptfour}{{\color{gpt4color} \texttt{GPT4}}}

\title{Strategic Interactions between Large Language Models-based Agents in Beauty Contests}
\author{%
  Siting Estee Lu\thanks{\tiny{I am immensely grateful to my co-authors of another paper, \cite{guo2024economics}, Shangmin Guo, Haoran Bu, Haochuan Wang, Yi Ren, Dianbo Sui, Yuming Shang, for providing some of the experimental results that enabled the progression of this line of research. 
  I am also thankful to Ed Hopkins, Marco Pelliccia, Vincent Crawford, and audiences at European Economic Review Summer School, for their valuable suggestions.}}\\
  School of Economics\\
  University of Edinburgh\\
  \texttt{s.lu@ed.ac.uk} \\
}

\begin{document}
\maketitle

\begin{abstract}
    The growing adoption of large language models (LLMs) presents potential for deeper understanding of human behaviours within game theory frameworks.
    Addressing research gap on multi-player competitive games, this paper examines the strategic interactions among multiple types of LLM-based agents in a classical beauty contest game.
    LLM-based agents demonstrate varying depth of reasoning that fall within a range of level-$0$ to $1$, which are lower than experimental results conducted with human subjects, but they do display similar convergence pattern towards Nash Equilibrium (NE) choice in repeated setting.
    Further, through variation in group composition of agent types, I found environment with lower strategic uncertainty enhances convergence for LLM-based agents, and having a mixed environment comprises of LLM-based agents of differing strategic levels accelerates convergence for all.
    Higher average payoffs for the more intelligent agents are usually observed, albeit at the expense of less intelligent agents.
    The results from game play with simulated agents not only convey insights on potential human behaviours under specified experimental set-ups, they also offer valuable understanding of strategic interactions among algorithms.
\end{abstract}
%\small{\textbf{JEL:}} C70, C60, C90
%\\
%\small{\textbf{Keywords:} Large Language Models, Beauty Contests, Strategic Interactions}

\section{Introduction}
\label{sec:intro}

With the emergent line of research surrounding the study of large language models (LLMs) capabilities, there is also growing discussions over the implications of LLMs on economic research and social sciences experiments, particularly in the field of game theory.
One of this work's main objectives is to make a case for using LLMs as simulated agents in economic games to shine a light on potential strategic behaviours.
Since LLMs are trained based on human-generated data, observing interactions between them could be fairly relatable to human subjects in experiments, and offer more insights than conventional simulation methods.
As opposed to diving into more expensive human-based experiments straightaway, it is also relatively easy and cost-effective to toy with different set-ups prior to concentrating on designs that are worth pursuing.

Recent research mainly focused on exploring 2-player cooperative and non-cooperative games, and often consists of a single LLM type (\cite{horton2023large}, \cite{phelps2023investigating}, (\cite{akata2023playing}).
While they provide interesting baselines for evaluating strategic behaviours, assuming agent homogeneity could make behaviour modelling more restrictive and does not leverage on the potential of having multiple LLM types.
Furthermore, competitive games involve more strategic consideration in predicting and attempting to outmaneuver opponents, exploring such games could offer new insights on strategic interactions that is different from the other games, and could provide novel use cases for LLMs that are worth exploring.
In this paper, I investigate a classical multi-player competitive game widely studied in Economics -- beauty contests.
Under this framework, agents' strategic levels and adaptive learning behaviours are jointly explored.
I found that LLM-based agents have strategic levels in between $0$ and $1$, evaluated using \cite{nagel1995unraveling}'s level-k model.
In repeated setting with revelation of past information, most of them show convergence towards the Nash equilibrium (NE) choice.
Since opponent types could be important contributing factor to adaptive learning, variations in proportion of agents types within the group and their impact on game outcomes are investigated.
When facing fixed-strategy opponents, LLM-based agents display greater convergence tendency in low strategic uncertainty environment; %The more intelligent LLM-based agents show incremental adjustments in choices as compared to abrupt change shown by the less intelligent agents when facing fixed-strategy opponents.
When the more intelligent LLM-based agent are playing against less intelligent agents, both agent types could show faster learning rate than when they are playing against their own types.
These results contribute first to assessing the models with a human-based metric on strategic levels, thereby drawing relation to represent heterogeneous human subjects with different LLM types, and via simulation of various set-ups, one can explore the potential strategic behaviours and postulate the possibility of inducing faster learning by varying group composition of agent types.

On a broader view, this work not only hope to highlight the plausibility of using LLM-based agents as a tool for social science research, the theories that were developed to explain and evaluate human behaviours can unequivocally help us to understand how this new era of computer algorithms would behave when competing against each other.
With the growing integration of LLMs into daily life, LLMs could be used as surrogate agents to communicate and interact with one another.
Understanding how algorithms interact could have significant social implications and diverse applications.

\section{Background}
\label{sec:background}

\textbf{LLMs as computational model of human behaviour:} Since the training process of LLMs rides on top of human-generated data and refinements based on direct human feedback, human reasoning process are baked into the algorithms, therefore, it is proposed that LLMs can be perceived as implicit computational model of human behaviour.
(\cite{ouyang2022training}, \cite{openai}, \cite{horton2023large})
Herein, I hope to streamline and differentiate between the two main aspects of how LLMs' human-like behaviour could cater to research for the social sciences community:

\textit{(a) Imitation of decision-making with known constraints.}
This approach uses LLMs to create synthetic agents with given profiles or constraints, and its objective is more grounded in granulating the elements contributing to decision-making.
It resembles agent-based modelling (ABM), where agents are pre-programmed to behave as we expect, and the outcome from which serves as a form of visualization and checkpoint of the theoretical predictions.
In beauty contests, this implies setting the strategic levels of the LLM-based agents a priori and examine their behaviours in comparison to theoretical predictions of agents of a certain level.

\textit{(b) Mirroring human-like behaviours without known constraints.}
By abstracting away from putting restrictions on behaviours a priori, simulation conducted with LLM-based agents essentially offers a tool for computational experiments.
In beauty contests, this approach implies identifying the intrinsic strategic levels of the LLM-based agents under pre-specified temperature.
Their behaviours with different experimental designs can be used to speculate the possible outcomes that could be anticipated if conducted with human subjects.
The extra benefit of this is that there is more flexibility in endogenous changes of strategic levels over time in repeated games.

\textbf{LLMs as heterogeneous agents.}
There are many ways to define agent heterogeneity.
One of which could be on the basis of differences in the underlying training data.
For instance, Anthropic's reward model training data primarily comes from crowd-sourcing feedback through Amazon Mechanical Turk, a platform often used for social sciences research;
and OpenAI's models are mainly trained on used prompts. (\cite{huggingface})
LLMs could also comprise of different priors and come in varying sizes, leading to different performances in text-based generating ability, thereby making them heterogeneous agents.
However, while the above distinctions of types are straightforward, it does not necessarily imply heterogeneity in strategic situations that I seek to study.
Therefore, in this paper, LLM types are characterized by their strategic levels, determined during the one-shot beauty contests, using a measure ubiquitous to how we evaluate the strategic types of human subjects and drawing parallels between the two.
%Since size of models could matter for strategic levels, this paper also builds a case where smaller language models can be useful in helping us to understand human behaviour, rather than just learning from larger models.

\textbf{LLMs as complements to human participants.}
At the core of discussions surrounding the usefulness of LLMs in social sciences research, there exists an important question of whether they can rise up to the task of participating in social experiments in place of human subjects or rational players.
There are growing replications of social experiments and strategic games to investigate this question, and while it was found that LLM-based agents deviate away from game-theoretical predictions and may be far from rational, they inevitably demonstrate ability to imitate human behaviours, making them human-like participants. (\cite{argyle2023out}, \cite{webb2023emergent}, \cite{huijzer_hill_2023}, \cite{dillion2023can}, \cite{guo2023gpt}, \cite{aher2023using}, \cite{mei2024turing}, \cite{fan2023can}, \cite{guo2024economics})
%Firstly, current literature have explored if LLMs, mostly a single type (often ~\gptthree~ or ~\gptfour~), could imitate human behaviours by replicating social experiments, and they found LLMs to be good proxies for aggregate level human cognition, and display strong capability in analogical reasoning, comprehension and communication skills to solve problems, as well as producing moral judgements that are well-aligned with human subjects. (\cite{argyle2023out}, \cite{webb2023emergent}, \cite{huijzer_hill_2023}, \cite{dillion2023can})
%Findings for many strategic games, such as the ultimatum game, wisdom of crowds, were also reproduced, where they were found to exhibit behavioral and personality traits like risk-aversion and cooperation, as well as learning patterns, similar to that of many human subjects.
%(\cite{guo2023gpt}, \cite{aher2023using}, \cite{mei2024turing})
%There are also works investigating how LLMs compare to rational players in games, and found them to deviate away from game-theoretic predictions.
%(\cite{fan2023can}, \cite{guo2024economics})
The main concern about using LLM-based agents is the opacity of their minds. (\cite{dillion2023can})
Although the same can be said about human minds, it is argued that there exist many theories to describe human reasoning in strategic situations, but a lack of equivalent to decipher the ``thinking" process of AI algorithms.
However, since LLMs are trained on human-generated data, which includes reasoning procedures, they could develop mechanisms similar to that of human brain, thus theories applied to human might also be applicable for explaining LLMs' behaviours.
(\cite{kosinski2023theory})
Despite this connection, it is still important to treat simulated results with care, thus my work does not aim at arguing for replacing human subjects in experiments with LLM-based agents completely, but rather using them as complements to shed some light on potential strategic behaviours.

\textbf{Choice of beauty contests.}
I focused on beauty contest games, which provide a desirable set-up that encompasses both competitive nature and interactions between multiple, and possibly heterogeneous, agents, whose level of reasoning can be easily distinguished.
%via iterated elimination of weakly dominated strategies, level-k or cognitive hierarchy models. 
(\cite{nagel1995unraveling}, \cite{camerer2004cognitive})
The game can also be constructed with a single interior NE solution, even in repeated setting, obstructing away from the complication of analyzing multiple equilibria.
Furthermore, there are many applications of beauty contest games with substantial social value.
For instance, the Keynesian Beauty Contest started off with the practical application to describe the stock market.
%where investors form expectations about others' strategies.
(\cite{keynes1936general}, \cite{nagel2017inspired})
With the market becoming more computerized, crypto trading bots emerge and function by executing pre-defined buying and selling strategies. (\cite{trality})
The backbone of these automatic bots can be replaced in the future by LLMs that account for vast human data on trading behaviours, and one could instead be focusing on choosing between proxies backed by different LLMs.
Therefore, understanding LLM-interactions could better inform us about the potential social implications, and beauty contests is a good starting point.
%\textbf{Strategic interactions between machines.} For centuries, economists, psychologists and neuroscientists have been studying the quasi-black box of human brain and came up with numerous theories that seek to explain human decision-making, they can be used to potentially explain machine's behaviour as well, or at least as a first step in connecting the two.
%Unlike previous generations of static computer algorithms that execute pre-programmed strategies, and behave regardless of the types of opponents they are faced with, LLMs are more human-like and dynamic in nature.
%\cite{kosinski2023theory} has explored the application of theory of mind for LLMs and \cite{dillion2023can} also mentioned comparing LLMs with human judgements could perhaps teach us about the machine minds of LLMs.
%By using concepts borrowed from human-based research, we can better understand how machines function in strategic setting.
%This is a step forward in improving their future performance to be more human-like or even, exceeding average human capacities.

\section{Beauty Contest Games}
\label{sec:beautycontests}

In this section, I first explore the one-shot and repeated beauty contests involving multiple LLMs:
~\chatglmtwo~, ~\chatglmthree~, ~\llama~, ~\baichuan~, ~\claudeone~, ~\claudetwo~, ~\palm~, ~\gptthree~, ~\gptfour~.
The results in my analysis are based on experimental data adapted from \cite{guo2024economics}.
Different from \cite{guo2024economics}, whose main objective is to evaluate LLMs' performance relative to the rational players that select NE choice, this work aims to analyze LLMs' behaviour as though they are human players.
Following which, I then choose two types of LLMs to construct groups of heterogeneous agents, and analyze how variations in group composition could affect learning pattern.
The additional computing resources required to conduct the additional set-ups are not substantial, but could have profound social significance.

\textbf{General Experimental Design.} Using a modified set-up following \cite{nagel1995unraveling}, and an exemplary prompt following \cite{guo2024economics} (recited in Appendix~\ref{originalprompt}\label{return_historical_prompt}):

Agents are asked to choose a number between $0$ and $\bar{c}$, where $\bar{c}$ is randomly generated from $0$ to $1000$.
One choosing closest to $p$, $p=\frac{2}{3}$, of the average wins the game.
A fixed prize of $\$x$ is awarded to the winner, and the prize is split amongst those who tie.
In repeated setting, the same game is played for $6$ periods, and agents are given historical information up to $3$ past periods, which include choices made by all agents, average of these choices, $\frac{2}{3}$ of the average, and past winners.
The limitation on revealing up to $3$ past periods is due to token restrictions to control computation intensity.
Therefore, this can be perceived as a partial feedback set-up or one with forgetting parameter.

\textbf{Analysis Focus.} The two main concepts central to the analysis are:
\begin{enumerate}
    \item \textit{Determination of Strategic Levels.} 
    Following \cite{nagel1995unraveling}, an agent is of strategic degree $n$ if he chooses a number $r(\frac{2}{3})^n$, where $r$ is defined to be the reference point, characterized by naive player or a point of salience in heuristics.
    In one-shot games and in period 1 of repeated games, this is assumed to be the mean of the range of numbers that can be chosen.
    \item \textit{Convergence.} In repeated setting, changes in choices are tracked to determine if there is convergence to the unique NE of $0$.
    The convergence rate is computed as $c_{t}=\frac{-(a_{t+1}-a_{t})}{a_{t}}$, where $a_{t+1} \leq a_{t}$, $a_t$ being the action/number chosen in period $t$.
    Changes in strategic levels are found by re-adjusting the reference point to the mean of the previous period choices.
    %\item \textit{Evolution of Strategic Levels.}
    %Strategic levels across period are found by re-adjusting the reference point at the beginning of each period to the mean of the previous period.
    %If strategic level, $n$, increases over time, then there is increasing level of thinking or revision in beliefs that opponents could be of higher strategic levels.
    %Tracking across periods also provides an average strategic level that does not simply depend on round 1 choices.
    %\item \textit{Evolution of Payoffs.}
    %Exploration of payoffs and their transition over time could determine how agents perform, depending both on their own strength as well as the environment.
\end{enumerate}

\textbf{Data collection.}
The experiments are conducted with API calls of different LLMs, providing a collection of independent observations that allows for a robust measure of strategic level for each LLM type.
In repeated settings, the information availability can be explicitly controlled through prompts that reveal histories perfectly or selectively to LLMs.
(\cite{bauer2023decoding})
While the stochasticity of model responses is dependent on the temperature selected, \cite{chen2023emergence} shows that strategic or choice consistency is less influenced by temperature, which depends more on the underlying reasoning process.
Therefore, the set-ups used in this paper use the default temperature.

%\vspace{-\baselineskip}
\begin{wrapfigure}{l}{0.5\textwidth}
    \centering
    \includegraphics[width=0.45\textwidth, height=0.5\textheight]{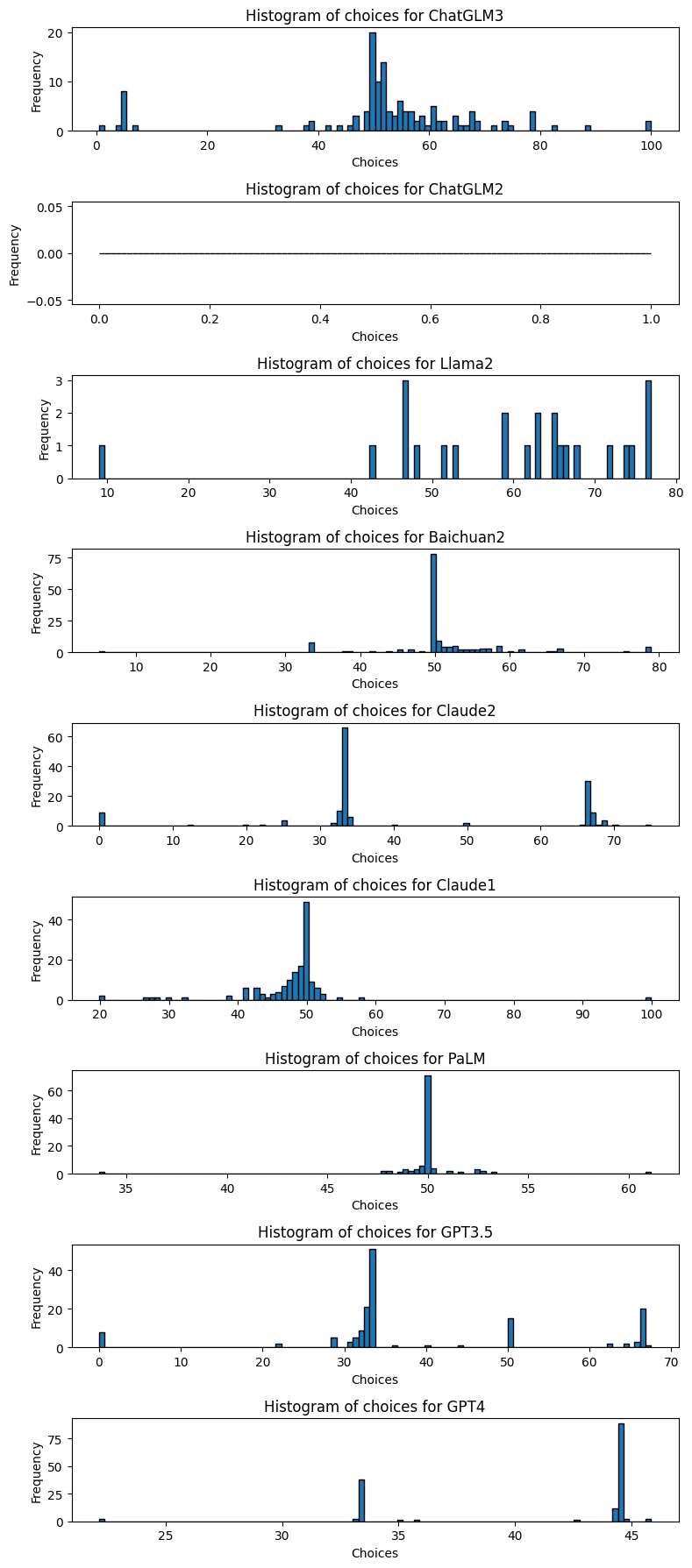}
    \caption{Frequency of choices}
    \label{fig:results:normalize_choices}
\end{wrapfigure}
%\vspace{-\baselineskip}

\textbf{One-Shot Game.}
150 sessions were ran with 9 agents back-boned by different LLMs.
In classical beauty contests, $\bar{c}$ is often fixed at $100$, all choices between $(66.66, 100]$ are weakly dominated by $66.66$, and those above $44.44$ are weakly dominated by $44.44$, etc.
Via iterative elimination of weakly dominated strategies, the number of steps taken determines agents' strategic levels.
Otherwise, going by level-k model with a focal point set at the mean of the number range, $50$, level-$0$ would choose $50$, and level-$1$ responds by choosing $33.33$, etc.
The unique interior NE solution of the game is $0$.
In this modified set-up with a randomly generated upper-bound for each game, the steps of assessing the strategic levels are unaffected.
For instance, using the level-k model, level-$0$ would simply choose the focal point, $\frac{\bar{c}}{2}$, and level-$1$ would respond by choosing $\frac{2}{3}\frac{\bar{c}}{2}$.
%The average strategic levels of each model are computed in this manner to ensure a robust and consistent measure.

\textit{Choices.} As shown by Figure~\ref{fig:results:normalize_choices}, the normalized choices are concentrated on $50$ for ~\chatglmthree~, ~\baichuan~, ~\claudeone~, ~\palm~.
As per level-k model, they are level-$0$ players.
~\llama~ records fairly dispersed and randomized choices,
%with an average around $60$, 
and thus can be perceived as level-$0$ as well.
~\claudetwo~ shows a spike around $33$, indicating the likelihood of level-$1$ thinking.
There is also a high choice frequency around $66$, which could be rationalized as step-1 of the iterated elimination of dominated strategies. (\cite{mauersberger2018levels})
For ~\gptthree~, most of the choices are concentrated around $33$, stipulating level-$1$ reasoning.
While there are some other spikes at $50$ and $66$, those are of much lower frequency.
%Further, for ~\claudetwo~ and ~\gptthree~, they chose the NE choice of $0$ $6\%$ and $5.3\%$ of the sessions respectively.
~\gptfour~ displays the highest spike in choices around $44$, going by iterated elimination of dominated strategies, this shows step-$2$ depth of reasoning.
A lower frequency spike is also observed around $33$, indicating level-$1$ thinking under the level-k model.
This could imply ~\gptfour~ has a level in between $1$ and $2$.
There are no data for ~\chatglmtwo~, indicating it is unable to complete the games and produce comprehensible output given the instructions.
%Based on the choice frequency, the larger language models often generate higher frequency in choices that can be characterized as higher level of strategic thinking.

%\vspace{-\baselineskip}
\begin{table}[H]
  \centering
  \resizebox{\textwidth}{!}{
\centering
\begin{tabular}{llllllllll}
\hline
\multicolumn{1}{|l|}{\textbf{Models}} &
  \multicolumn{1}{l|}{~\chatglmthree~} &
  \multicolumn{1}{l|}{~\chatglmtwo~} &
  \multicolumn{1}{l|}{~\llama~} &
  \multicolumn{1}{l|}{~\baichuan~} &
  \multicolumn{1}{l|}{~\claudetwo~} &
  \multicolumn{1}{l|}{~\claudeone~} &
  \multicolumn{1}{l|}{~\palm~} &
  \multicolumn{1}{l|}{~\gptthree~} &
  \multicolumn{1}{l|}{~\gptfour~} \\ \hline
\multicolumn{1}{|l|}{\textbf{Average}} &
  %52.0292929992966
  \multicolumn{1}{l|}{52.029} &
  \multicolumn{1}{l|}{N/A} &
  %59.519266481657276
  \multicolumn{1}{l|}{59.519} &
  %51.15846805353594
  \multicolumn{1}{l|}{51.158} &
  %41.608918220216836
  \multicolumn{1}{l|}{41.609} &
  %47.69585017993402
  \multicolumn{1}{l|}{47.696} &
  %49.975701986927916
  \multicolumn{1}{l|}{49.976} &
  %38.91214603534093
  \multicolumn{1}{l|}{38.912} &
  %41.0723539994837
  \multicolumn{1}{l|}{41.072} \\
\multicolumn{1}{|l|}{\textbf{Median}} &
  %51.724137931034484
  \multicolumn{1}{l|}{51.724} &
  \multicolumn{1}{l|}{N/A} &
  %62.685269131995895
  \multicolumn{1}{l|}{62.685} &
  \multicolumn{1}{l|}{50.0} &
  %33.33333333333333
  \multicolumn{1}{l|}{33.333} &
  %49.31300583474497
  \multicolumn{1}{l|}{49.313} &
  \multicolumn{1}{l|}{50.0} &
  %33.33333333333333
  \multicolumn{1}{l|}{33.333} &
  %44.44228199822643
  \multicolumn{1}{l|}{44.442}
  \\ \hline
  \end{tabular}
}
  \caption{Average and Median Choice of the LLMs across 150 Sessions}
  \label{tab:average_median_choice}
\end{table}
\vspace{-\baselineskip}

\cite{nagel1995unraveling} and \cite{bosch2002one} have conducted beauty contests with different human populations, such as students (mean=$36.73$, median=$33$), theorist (mean=$17.15$, median=$15^*$), newspaper readers (mean=$23.08$, median=$22^*$), etc.\footnote{\footnotesize{The median with $*$ are guesstimated based on the figures in \cite{nagel1995unraveling}, \cite{bosch2002one}.}}
Human subjects show strong deviation away from game-theoretic prediction, and display on average iteration steps $1$ and $2$ evaluated by the level-k model.
%Even though it is often expected that the general public is more noisy in choices, in this case, newspaper readers have slightly lower mean than the student population, possibly due to longer time for reflection and thereby making more contemplated decisions. (\cite{mauersberger2018levels})
Compared to them, LLM-based agents are choosing slightly higher numbers, as shown in Table~\ref{tab:average_median_choice}, which corresponds to an average strategic level between $0$ to $1$.
This result comply with the impression that general public could display more randomized choices, and each LLM has a different strategic level, which can represent different human subjects or subsets of population.
%As a result, a single LLM-based agent could be more representative of the general population instead of a specific type of individuals.
%%Given the alignment of models are based on data potentially generated by different groups of labelers and users, agents back-boned by different LLMs can represent different subsets of the populations.
%the different average numbers selected by the models could also be used to imply behaviours of specific groups of human subjects that choose different numbers on average.
\\ \ \\
For human subjects, when given identical game set-up, it is possible that they might employ different strategies. (\cite{devetag2016eye}, \cite{costa2008stated})
The same could apply to LLM-based agents.
It could be important to understand how varied one's choice might be given the same instructions.
By fixing the upper-bounds to the same value, ~\claudetwo~, ~\gptthree~ and ~\gptfour~ displayed more variability in choices as compared to other models
(Appendix~\ref{choicevariability}\label{choicevariabilityreturn})
This shows that choices might not be static even when the instructions are exactly the same.
While \cite{bauer2023decoding} indicates running multiple sessions could already accommodate the stochastic nature of LLM responses, my method of computing for average strategic levels further account for both identical and different upper-bounds, rendering a more robust and consistent measure for each model.
%The determination of average choices and the corresponding strategic levels based on many sessions of one-shot games that encompass both identical upper-bounds and different upper-bounds would therefore render a more consistent and robust measure.

\begin{figure}[H]
     \centering
    % \begin{subfigure}[b]{0.42\textwidth}
    %     \centering
    %     \includegraphics[width=\textwidth]{graphs/fig1-mixed-one-shot.png}
     %    \caption{10 Sessions}
     %    \label{fig:mixed-one-shot1}
     %\end{subfigure}
     %~
     \begin{subfigure}[b]{0.42\textwidth}
         \centering
         \includegraphics[width=\textwidth]{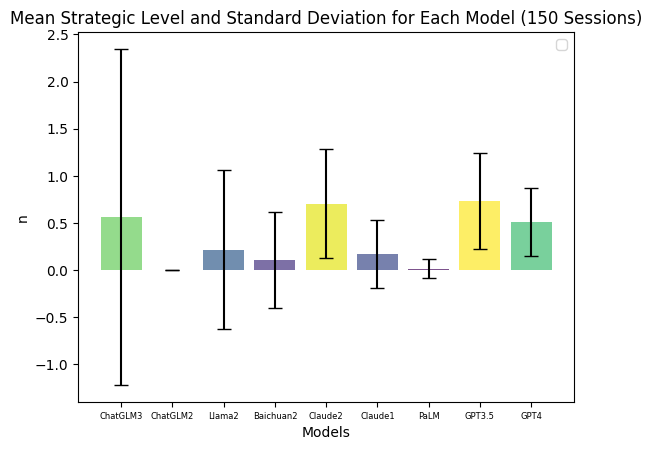}
         %{graphs/mean_strategic_level_reference_average.png}
         \caption{$r=\frac{\bar{c}}{2}$}
        \label{fig:mixed-one-shot2}
     \end{subfigure}
     ~
     \begin{subfigure}[b]{0.42\textwidth}
         \centering
         \includegraphics[width=\textwidth]{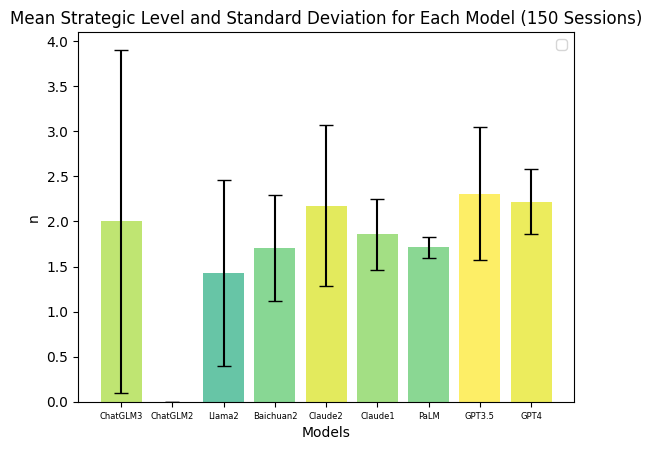}
         %{graphs/fig2-mixed-one-shot.png}
         \caption{$r=\bar{c}$}
        \label{fig:mixed-one-shot1}
     \end{subfigure}
     \caption{Average strategic levels of LLM-based agents with different reference points.}
     \label{fig:results:mixed-one-shot}
\end{figure}

\textit{Reference dependence.}
The reference point of level-k model is defined to be the choice of a non-strategic agent, and is assumed to be the mean of the number range, pertaining to insufficient reasoning. (\cite{mauersberger2018levels})
However, this focal point can be disputable.
In my set-up, the varied upper-bounds may well be the focal points rather than taking the extra step of computing for the mean.
Figure~\ref{fig:results:mixed-one-shot} shows the average strategic levels are between $0$ and $1$ given the reference point $r=\frac{\bar{c}}{2}$, and between $1$ to $2.5$ when it is $r=\bar{c}$.
In the following sections, I evaluate the results using the conventional focal point of $\frac{\bar{c}}{2}$.
Figure~\ref{fig:mixed-one-shot2} shows the strategic level to be high for ~\chatglmthree~, ~\claudetwo~, ~\gptthree~ and ~\gptfour~.
Surprisingly, ~\gptfour~ has slightly lower strategic level than ~\gptthree~, even though it is often presumed to be a stronger model.
Its lower depth of reasoning could be due to its capability or it being trained on more data, thereby encompassing higher possibility of noisy strategies that leads to it choosing higher numbers.
%``belief" that its opponents have relatively low strategic levels.
%%Further, the variability in levels could also differ, suggesting models can represent agents of differing strategic level, or even variability in depth of reasoning.

%\vspace{-\baselineskip}

\vspace{-\baselineskip}
\begin{wrapfigure}{r}{0.45\textwidth}
\vspace{-\baselineskip}
    \centering
    \includegraphics[width=0.4\textwidth, height=0.16\textheight]{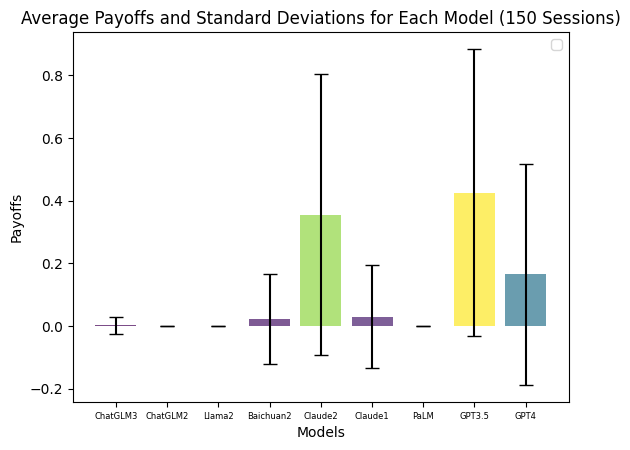}%{graphs/mean_payoffs.png}
    \caption{Average payoffs.
    %Slightly larger variability for stronger models.
    }
    \label{fig:results:mean_payoffs}
\end{wrapfigure}

\textit{Payoff.}
%Having high strategic level does not necessarily imply that the LLM-based agent will win the games and obtain the high payoffs.
Figure~\ref{fig:results:mean_payoffs} demonstrates that ~\claudetwo~, ~\gptthree~ and ~\gptfour~ have relatively higher average payoffs than the others, out of which, ~\gptthree~ have the highest average payoffs as compared to the other models.
%However, ~\gptthree~ also has relatively high standard deviation, which indicates considerable variability in payoffs.
Associating the results with strategic levels, LLM-based agents with higher average strategic levels often obtain higher average payoffs, except for ~\chatglmthree~, which could be due to high variability in its strategic levels that might have adversely influenced its average gain.
\begin{wrapfigure}{l}{0.5\textwidth}
    \centering
    \includegraphics[width=0.41\textwidth, height=0.17\textheight]{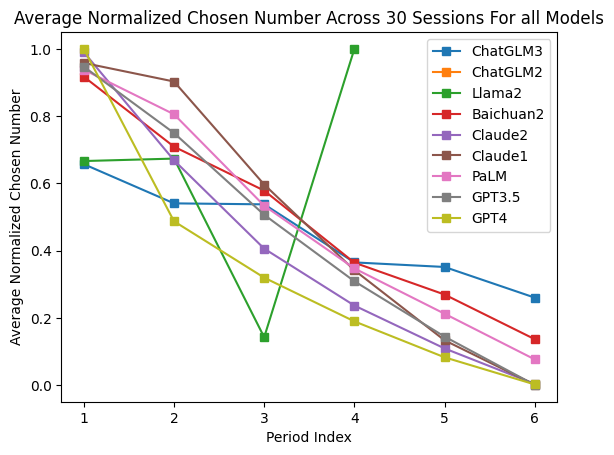}
    \caption{Convergence in average chosen number.}
    \label{fig:results:chosen_number_multiperiod}
\end{wrapfigure}

\textbf{Repeated Games.} $30$ sessions of repeated beauty contests were ran, following the repeated set-up highlighted in the general experimental design.
In Figure~\ref{fig:results:chosen_number_multiperiod}, most LLM-based agents show convergence in actions, particularly for ~\claudeone~, ~\claudetwo~, ~\gptthree~, and ~\gptfour~, which are the models with higher strategic levels.
Their chosen numbers are approximately $0$ in period $6$, indicative of them learning to play NE choice across time.
%(A sample of convergence paths for each LLM-based agents over the $6$ periods and for $10$ sessions is illustrated in Appendix~\ref{convergence_multillm}, where the more intelligent agents show higher chances of convergence.)
%chatglm3-A, chatglm2-B, llama2-C, baichuan2-D, claude2-E, claude1-F, palm-G, gpt3-H, gpt4-I
%\begin{figure}[H]
%     \centering
%     \begin{subfigure}[b]{0.49\textwidth}
%         \centering
%         \includegraphicsc[width=\textwidth]{graphs/period1_average_n_c.png}
         %{graphs/period1_n.png}
%         \caption{Period 1 strategic level}
%        \label{fig:results:period1_average_n}
%     \end{subfigure}
%     ~
%     \begin{subfigure}[b]{0.47\textwidth}
%         \centering
         %\includegraphics[width=\textwidth]{graphs/normalized_chosen_number(30).png}
%         \includegraphics[width=\textwidth]{graphs/average_normalized_chosen_number.png}
%         \caption{Average chosen number}
%        \label{fig:results:chosen_number_multiperiod}
%     \end{subfigure}
%     \caption{Strategic level determined by period 1 choice and convergence in average chosen number over 6 periods.}
%     \label{fig:results:multi_periods}
%\end{figure}

\begin{figure}[H]
     \centering
     \begin{subfigure}[b]{0.46\textwidth}
         \centering
         \includegraphics[width=\textwidth]{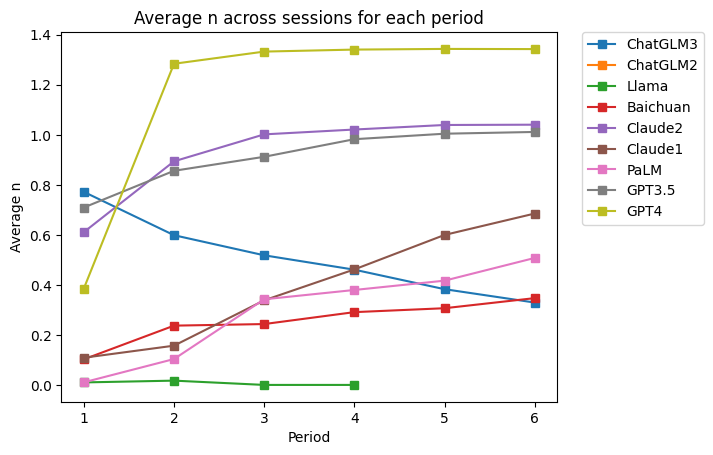}
         %{graphs/period1_n.png}
         \caption{Averaging strategic levels}
        \label{fig:results:average_n_across_sessions}
     \end{subfigure}
     ~
     \begin{subfigure}[b]{0.46\textwidth}
         \centering
         \includegraphics[width=\textwidth]{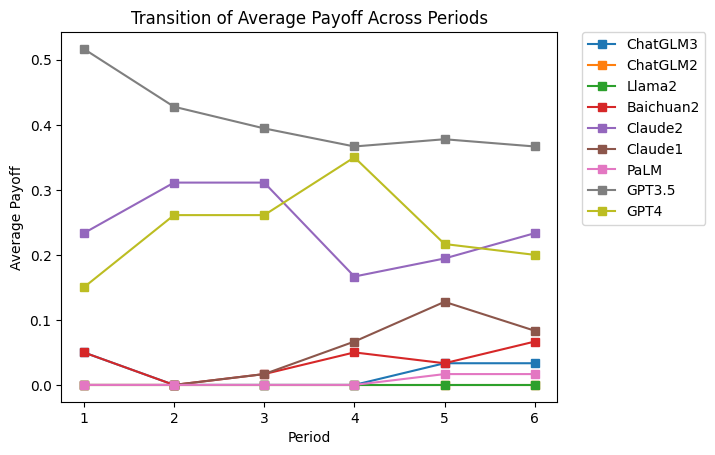}
         \caption{Average Payoffs}
        \label{fig:results:average_payoffs_across_sessions}
     \end{subfigure}
     \caption{Averaging strategic levels and average payoffs across 30 sessions for 6 periods.}
     \label{fig:results:multi_periods2}
\end{figure}

\textit{Frequency and Evolution of Choices and Strategic Levels:}
I explore the changes in strategic level for each model, averaged across sessions, for each period in Figure~\ref{fig:results:average_n_across_sessions}.
The strategic levels evolve over time, but the range of which is rather narrow, and on average, they stay within the bound of $0$ and $1.4$.
Most LLM-based agents shows increasing depth of reasoning, especially ~\claudetwo~, ~\gptthree~ and ~\gptfour~.
An interesting observation is that while ~\gptthree~ has higher strategic level than ~\gptfour~ in one-shot games, in repeated setting, ~\gptfour~'s average strategic level surpasses that of ~\gptthree~ from periods 2 onwards, implying that it could be more adept at revising its beliefs about opponents over time given past information.
The abnormality in Figure~\ref{fig:results:average_n_across_sessions} comes from ~\chatglmthree~ and ~\llama~, the first shows a decrease in average strategic level, indicating lack of ability to respond to historical information and adjust behaviour accordingly; the second display naive, random selection.
%(Further details in Appendix~\ref{freqency_choice_strategiclevel_repeated}\label{freqency_choice_strategiclevel_repeated_return})
%or that due to the high variance in agent's actions, contributing to a decrease in the average strategic level over time for this sample.

\textit{Payoff Evolution:} Figure~\ref{fig:results:average_payoffs_across_sessions} shows ~\gptthree~ outperforms the rest in all periods, while ~\claudetwo~ and ~\gptfour~ are more or less comparable.
The rest of the LLM-based agents do not obtain average payoffs as high, but most of them display growth over time.
Coupled with Figure~\ref{fig:results:chosen_number_multiperiod} that shows convergence in average choice towards NE, the increasing payoffs could be an indication of learning about the optimal action to take to win the game.
%and also obtain higher payoffs.
%The exception is ~\llama~, it does not finish the 6 periods and average actions display no convergence, thus its corresponding average payoffs is also consistently $0$ throughout.
%\vspace{-\baselineskip}
\vspace{-\baselineskip}
In this section, the purpose of one-shot games is to evaluate the strategic levels of LLM-based agents that I can draw parallel to human subjects.
The results from which resembles experiments conducted with human subjects, who show strong deviation away from game-theoretic prediction and tend to display low levels of reasoning.
However, the distinction is that LLM-based agents display even lower level of reasoning than that of human subjects in current literature.
The repeated setting shine a light on how the simulated agents could behave over time.
Similar to human subjects, LLM-based agents do not display iteration steps that go over $2$ within the span of the games, and they do seem to learn from historical information and show convergence towards NE choice.

\section{Adaptive Learning with Variation in Group Composition}\label{adaptivelearning}

In this section, I further analyze the strategic interactions between LLM-based agents by exploring their adaptive learning behaviour given variation in group composition.
I selected two LLM types, ~\gptthree~ and ~\palm.
Based on previous section, ~\gptthree~ has a strategic level of approximately $1$ and ~\palm~ has level-$0$, representing higher and lower intelligence agent type respectively, where intelligence is interpreted loosely as metonym for strategic level.
Herein, games are played among $10$ agents, who are asked to choose a number between $[0,100]$. Each game comprises of $5$ periods with full historical information disclosure.
(Details: Appendix~\ref{detailed_set_up}\label{convergence_llm_return})
%I explore two main environments: 
%\textit{Partial static environment.} A single LLM type will be playing against hard-coded fixed-strategy opponent(s);
%and \textit{Dynamic environment.}
%The interactions between the two LLM types will be explored.
%\textbf{Higher Intelligence Model} (represented by~\gptthree~) vs. \textbf{Lower Intelligence Model} (represented by ~\palm~) when playing against fixed-strategy opponents:

\textbf{Partial static environment: LLM vs. Static Algorithm.}
\label{llm_static}
In this setting, LLM-based agents are playing against fixed-strategy players, whose actions are hard-coded to be $0$.
Across different treatments, the proportion of fixed strategy players and LLM-based agents change, but the group size remains the same.
The 3 treatments are: (1) 1 LLM + 9 Hard-coded Agents (Low strategic uncertainty); (2) 5 LLMs + 5 Hard-coded Agents (Mixed strategic uncertainty); (3) 9 LLMs + 1 Hard-coded Agents (High strategic uncertainty).
An exemplary prompt is shown in Appendix \label{return_newprompt_fixedstrategy_NE} ~\ref{newprompt_fixedstrategy_NE}.
LLM-based agents are specifically told that some of their opponents are playing a fixed strategy of $0$.
%As the population may comprise a mixture of LLM-based agents and fixed strategy agents, by varying the proportion of fixed strategy players and LLM-based agents, the degree of strategic uncertainty varies across different sessions.
%The NE strategy would not be affected by the proportion of agent types, but it is expected that the speed of convergence towards NE could differ across settings, and would be faster when proportion of fixed strategy players is higher.
%

For both LLM types, there is convergence in choices to $0$, exhibiting either refinement of belief about opponents' strategies or progression in their depth of strategic thinking when given historical information.
The pace is slower as strategic uncertainty grows, where the proportion of LLM-based agents becomes larger relative to fixed strategy agents.
%but their results for set-up 1 and 2 largely coincide, indicating that lower intelligence agent(s) are not very sensitive to the difference in environments of having 90\% vs. 50\% fixed strategy opponents in the group.
%Alongside larger fluctuations in choices in set-up 3 that comprises of even lower proportion of fixed strategy opponents, this suggests that higher strategic uncertainty could induce greater variability in strategies and might lead to non-convergence behaviour in lower intelligence agents.
Comparing between the high and low types, when strategic uncertainty is high, low types display larger variability in behaviour and they might not converge at all.
Furthermore, low types are also less ``cautious" in a sense that they could converge to $0$ in period $2$ straightaway when strategic uncertainty is relatively low, while convergence to $0$ takes a gradual process for high types.
This could indicate that high types are going from less sophisticated strategies to more refined choices through iterative learning and adaptation, and the lack of such systematic adjustments in choices for the low types could suggest that they are relying more on intuitive guesses than successive elimination of less likely options.
%\vspace{-\baselineskip}
\begin{figure}[H]
     \centering
     \begin{subfigure}[b]{0.38\textwidth}
         \centering
         \includegraphics[width=\textwidth]{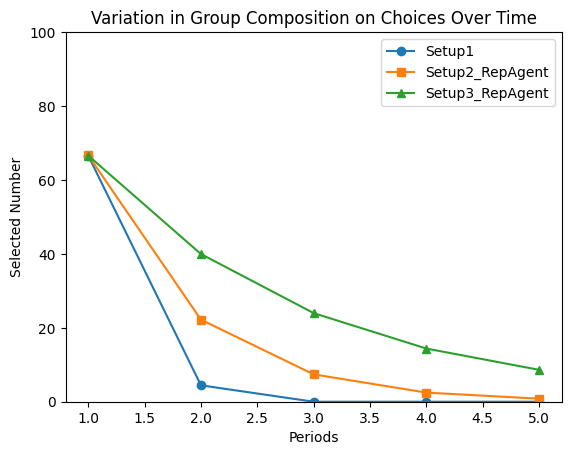}
         \caption{Higher Intelligence}
        \label{fig:results:choices_with_fixed_strategy_opponents_strong}
     \end{subfigure}
     ~
     \begin{subfigure}[b]{0.5\textwidth}
         \centering
         \includegraphics[width=\textwidth]{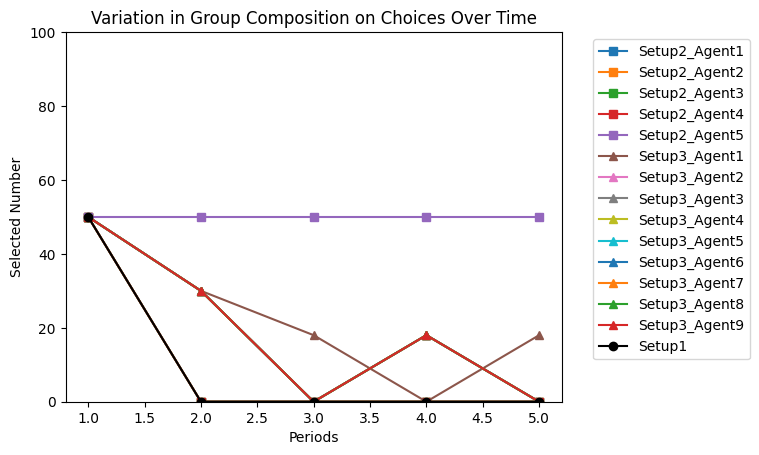}
         \caption{Lower Intelligence}
        \label{fig:results:choices_with_fixed_strategy_opponents_weak}
     \end{subfigure}
     \caption{Transition in choices of LLM-based agents playing against fixed strategy opponents.}
     \label{fig:results:choices_with_fixed_strategy_opponents}
\end{figure}
\vspace{-\baselineskip}
\textit{Convergence of Choices and Evolution of Strategic Levels.}
For both types, decreasing strategic uncertainty is related to higher convergence rate.
%While convergence for the low types could be faster than the high types due to an abrupt jump to the NE choice, larger noise is also recorded.
When evaluating the transition in strategic levels across periods, high types show transition from $0$ to $1$, 
%before playing the NE choice, 
while most of the low types stay at level-$0$, with some fluctuations between $0$ and $1$ when strategic uncertainty is high and actions are more noisy.

\textit{Payoffs.}
%Figure~\ref{fig:payoffs_setup1_static_strong} shows that in the environment with low strategic uncertainty, the single higher intelligence LLM-based agent starts off with payoff of $0$, but as it converges to the NE choice of $0$, the prize are shared among winners and the agent obtains positive payoff from period $3$ onwards.
%In the mixed environment (Figure~\ref{fig:payoffs_setup2_static_strong}), though the actions of LLM-based agents converge to $0$, they remain far from $\frac{2}{3}$ of the average than fixed strategy players, therefore, prizes are shared by fixed strategy players and LLM-based agents have a flat payoff of $0$.
%In the environment with high proportion of LLM-based agents (Figure~\ref{fig:payoffs_setup3_static_strong}), the reverse happens, since $\frac{2}{3}$ of average is relatively high, the prizes are shared by the LLM-based agents, and they have a flat payoff of $0.11$.
%In general, when LLM-based agents are playing in a repeated beauty contest game with fixed strategy opponents, they display convergence in actions towards the interior NE strategy of $0$ over time given historical information of past choices and payoffs are revealed.
%As I vary the proportion of LLM-based agents and fixed strategy opponents, the speed of convergence is usually slower when there is presence of other simulated agents, which contributes to higher strategic uncertainty.
The payoffs are in favor of LLM-based agents when strategic uncertainty is relatively high.
%as $\frac{2}{3}$ of average would be much higher than $0$.
High types could gain better payoffs under low and high strategic uncertainty as compared to mixed environment, where they receive a flat payoff of $0$ throughout the periods.
Comparing between the types, interestingly, payoffs achieved in all settings by the low types could be comparable or even higher than that of the high types, though the variations is also larger.
This indicates that higher strategic level does not necessarily imply higher payoffs when competing against fixed strategy opponents.
This results not only signifies the potential game play if human subjects are playing against opponents that naively adopt a fixed strategy of $0$, it could also illustrate a possible outcome if they are going against static computer algorithms executing a fixed NE strategy.
(Appendix~\ref{detailed_set_up}\label{convergence_llm_return})
%In Figure~\ref{fig:payoffs_setup1_static_weak}, the single lower intelligence LLM-based agent shows similar payoff pattern as that of higher intelligence agent.
%In the mixed environment (Figure~\ref{fig:payoffs_setup2_static_weak}), unlike the higher intelligence agents, majority of the lower intelligence agents have non-zero payoffs.
%Lastly, for the high strategic uncertainty environment (Figure~\ref{fig:payoffs_setup3_static_weak}), majority of the lower intelligence LLM-based agents achieve positive payoffs across the periods.
%However, the payoffs are not flat, this is contributed by a case of a specific LLM-based agent ``lagging behind", where it chose a number that was chosen by majority in the past period but not in the current period.

\textit{Application.}
One example of beauty contest applications is the Bertrand competition model. (\cite{mauersberger2018levels})
LLM-based and fixed strategy agents can be perceived as firms adopting different pricing strategies, with the objective to win over the market and maximize their profits.
%Suppose there are $10$ firms in the market, each of them have a marginal cost of $\gamma$.
%They set the price of the product at time $t$ to be $p_{it}=\frac{2}{3}\hat{E}_{it} \min{(\gamma, p_{1t}, p_{2t},..., p_{10t})}-\gamma$, where $\hat{E}_{it}$ is the subjective expectation of firm $i$ held at time $t$.
Fixed strategy firms could be playing the equilibrium action by setting the price equals to marginal cost, while LLM-based firms could be more dynamic and adjust their prices in each period.
%Based on the simulation results previously, having higher proportion of fixed strategy firms would drive the prices set by LLM-based firms down faster, and firms with higher strategic level would adjust the prices down step-by-step, and those with lower strategic level either adjust straightaway or they failed to adjust at all.
If there exist certain rigidity in the short run, such as production capacity constraints for the firms or limited response time for the consumers, then those who set higher prices would be able to gain higher profits.
In the long run, however, all factor inputs are flexible and consumers will not purchase from the firm that sells a homogeneous product but at a higher price than the equilibrium.
%, therefore, it is better for firms to converge to the equilibrium price in order to stay in the market.
As a result, high type firms could often achieve better outcome than low types in the short run, where they can earn a positive profit by converging gradually.
Even in the long run, the larger variance in pricing strategies for the low types as compared to high types, where they either failed to converge or display high volatility in prices, could adversely impact their profits.
%A possible example could be competition for research funding.
%This resembles a beauty contest game, in which agents hope to secure funding would propose less ambitious projects within the submitted pool of proposals in order to increase the probability of being accepted.
%This could be a race to the bottom where the degree of practicality is weighed higher, while revolutionary but risky ideas are been driven down.
%The fixed-strategy players could be seen as those that always play the equilibrium action of proposing less innovative and less risky projects, while the LLM-based players are those that respond dynamically to the environment.
%As a social planner, I might want to encourage breakthroughs and hope researchers would submit more revolutionary and high-impact projects, by cultivating an environment like set-up 3, I would be able to encourage higher number selection.
If firms outsource their pricing strategies to automated algorithms, the simulation could also be interpreted as competition between different algorithms.
While automated pricing has been widely discussed in literature, those back-boned by LLMs that could respond to changes in rivals' strategies by adjusting their own could spark fresh perspectives. (\cite{brown2023competition}, \cite{chen2016empirical})
%\\ \ \\
%\textbf{Line of Reasoning.} In the prompt I ask for the reasoning for choosing. I need to show a few example of answers, and in general what happens.

\textbf{Dynamic environment: LLM vs. LLM.}
\label{llm_llm}
In this setting, LLM-based agents are playing against each other (~\gptthree~ is denoted as $H$, and ~\palm~ as $L$ henceforth).
The 5 treatments are: (1) 10 $H$ LLMs; (2) 9 $H$ LLMs + 1 $L$ LLM; (3) 5 $H$ LLMs + 5 $L$ LLMs; (4) 1 $H$ LLM + 9 $L$ LLMs; (5) 10 $L$ LLMs.
I use the original prompt in Appendix~\label{return_historical_prompt_llm} \ref{historical_prompt}.
%For the pure environments, the rate of change in choices is expected to be the same for the high and low types.
%As for set-ups $2$ to $4$, if high types chose a smaller number than low types because they go through more iterations of reasoning, then high types are expected to proportionally lower their estimations less from time $t$ to $t+1$ compared to low types.
%There could mean slower rate of change for the high types than low types.
%On the other hand, if high types have strong beliefs that they are playing against opponents who will choose higher numbers while low types believe the other way around, then it is possible for $\frac{a_{Ht}}{a_{Lt}}>1$, then the inverse happens, low types are expected to proportionally lower their estimations less from time $t$ to $t+1$ compared to high types.
%This would mean faster rate of change in choices for high types as compared to than low types.
%(Greater details of the choice variation over time and the possible speed of convergence is illustrated in Appendix~\ref{choice_variations_llm}\label{choice_variations_llm_return}.)
%and also compared to pure environment with high types; while there is expected to be slower convergence for low types in the mixed environment, as compared to high types and also in regards to pure environment that only contain low types.
%
%\textbf{Higher Intelligence Model} (represented by~\gptthree~) vs. \textbf{Lower Intelligence Model} (represented by ~\palm~) when playing against each other:

\begin{figure}[H]
     \centering
     \begin{subfigure}[b]{0.31\textwidth}
         \centering
         \includegraphics[width=\textwidth]{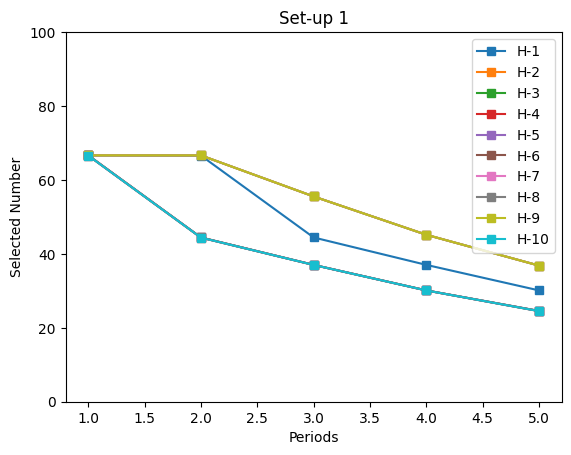}
         \caption{Pure High Intelligence}
         \label{fig:setup1_mixedllm}
     \end{subfigure}
     ~
     \begin{subfigure}[b]{0.31\textwidth}
         \centering
         \includegraphics[width=\textwidth]{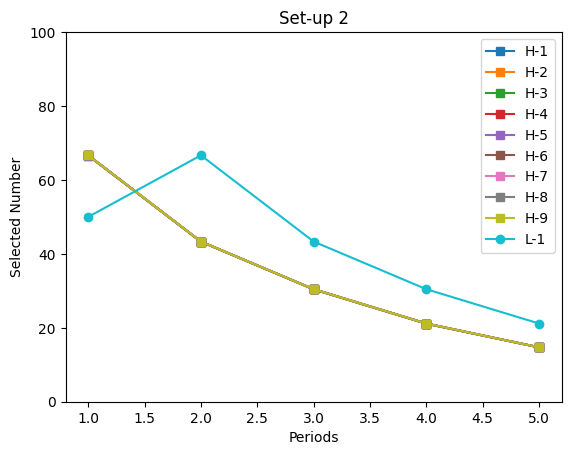}
         \caption{Highly Intelligent}
        \label{fig:setup2_mixedllm}
     \end{subfigure}
     ~
     \begin{subfigure}[b]{0.31\textwidth}
         \centering
         \includegraphics[width=\textwidth]{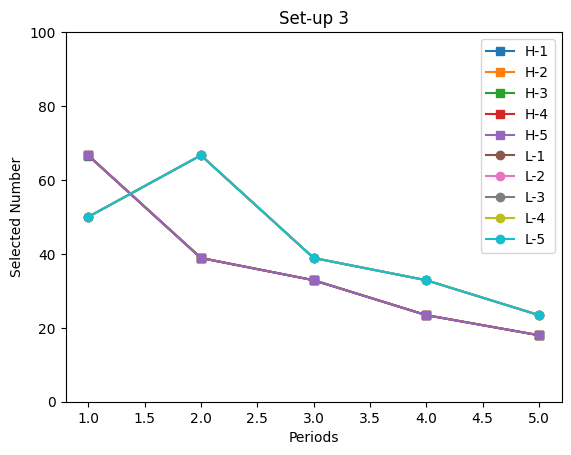}
         \caption{Mixed Intelligent}
         \label{fig:setup3_mixedllm}
     \end{subfigure}
     ~
     \begin{subfigure}[b]{0.31\textwidth}
         \centering
         \includegraphics[width=\textwidth]{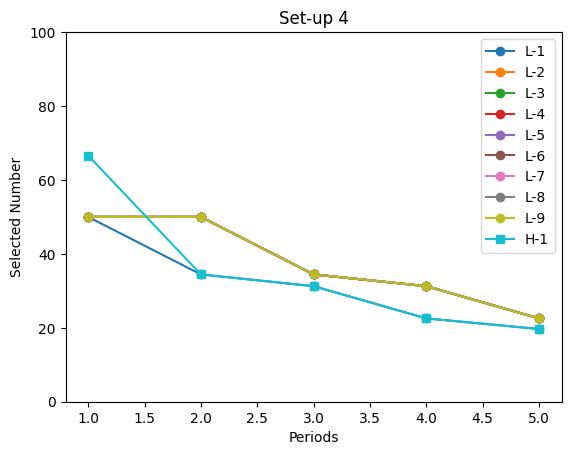}
         \caption{Less Intelligent}
        \label{fig:setup4_mixedllm}
     \end{subfigure}
     ~
     \begin{subfigure}[b]{0.31\textwidth}
         \centering
         \includegraphics[width=\textwidth]{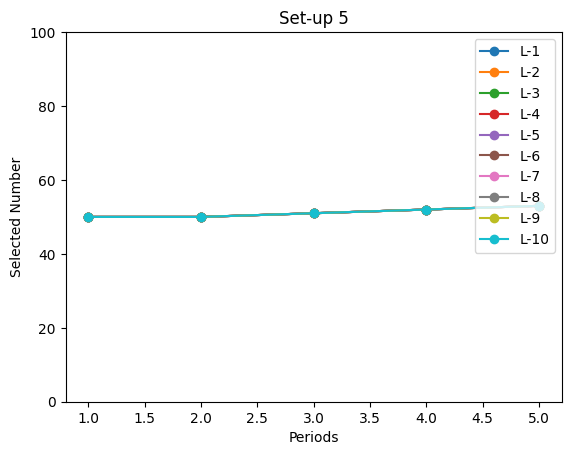}
         \caption{Pure Low Intelligence}
         \label{fig:setup5_mixedllm}
     \end{subfigure}
     \\
     \caption{Impact of variations in proportion of different LLM-based agents on chosen number.}
     \label{fig:chosen_number_mixedllm}
\end{figure}

In Figure~\ref{fig:chosen_number_mixedllm}, set-up 1 and 5 depict pure intelligence environments.
While the high types show adjustment in their choices to lower numbers, the low types persistently choose around $50$.
In set-up 2 to 4, both high and low type agents show convergence to lower numbers.
The main difference is that the gap between the numbers chosen by the high and low types is smaller when there is higher proportion of low types in the group.
%However, when they are placed in mixed environment, their learning is better facilitated when there exist high types in the group.
%High types, on the other hand, will respond to past plays regardless of the environment, but the variation in choices could be smaller when placed in the mixed environment.
The results shows that low type agents fail to adapt their strategies despite disclosure of historical information in the pure environment that only comprises of low types, but mixed environments could instigate faster learning, which applies for both high and low types, particularly when there is higher proportion of high types.
This put forth a strong statement that adding a single high type could very well stir the pot and speed up learning.
%\vspace{-\baselineskip}
\vspace{-\baselineskip}

\begin{wrapfigure}{r}{0.5\textwidth}
    \centering
    \includegraphics[width=0.5\textwidth, height=0.17\textheight]{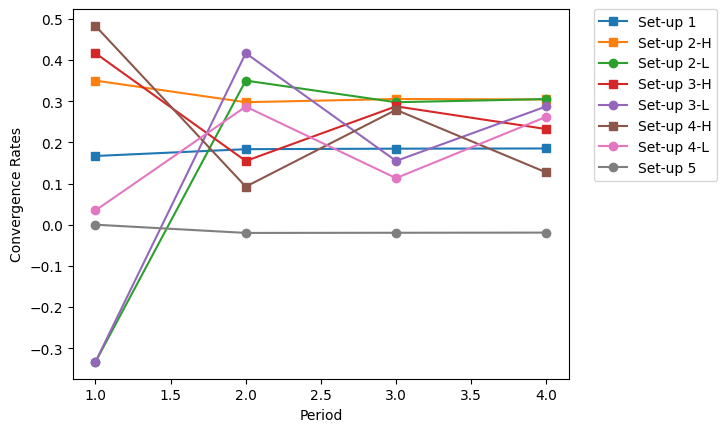}
    \caption{Average convergence rates.}
    \label{fig:results:convergence_rates_mixedllm}
\end{wrapfigure}

\textit{Convergence of Choices and Evolution of Strategic Levels:}
Figure~\ref{fig:results:convergence_rates_mixedllm} shows low and approximately flat convergence rates for set-up 1 and 5.
%and the learning rate in set-up 2-H is faster than pure environments.
In the mixed environments, the convergence speed fluctuates but could be higher than the pure environments.
For instance, most of the convergence rates in set-up 2 to 4 lay above the lines for set-up 1 and 5, and higher the proportion of $H$, the higher the convergence rates.
%as I increase the proportion of low types, the convergence speed of the high types fluctuates and is lower in general, but 
%%For instance, if the proportion of high types is 50\% or more, convergence rate could be higher for both types in the mixed environments than pure ones.
As for variations in strategic levels across time, all set-ups, except for 5 where low types do not display any apparent evidence of learning, shows changes in strategic levels.
In set-up 3 particularly, high types could reach a strategic level greater than $1$, which implies having highly mixed environment could also stimulate considerable growth in depth of reasoning for some agents.
A possible conjecture for this could be that the strategic landscape is more complex in a highly mixed environment, agents cannot simply default to strategies assuming similar reasoning process from all agents, and this induces increasing depth of reasoning.
%Similarly, for the less intelligent agents, having a mixed environment and higher proportion of low types, such as in set-up 3 and 4, is beneficial in instigating higher strategic levels.
%\begin{figure}[H]
%     \centering
%     \begin{subfigure}[b]{0.42\textwidth}
%         \centering
%         \includegraphics[width=\textwidth]{graphs/n_values_mixedllm1.png}
%         \caption{Higher Intelligence}
%        \label{fig:results:n_values_mixedllm_strong}
%     \end{subfigure}
%     ~
%     \begin{subfigure}[b]{0.42\textwidth}
%         \centering
         %\includegraphics[width=\textwidth]{graphs/normalized_chosen_number(30).png}
%         \includegraphics[width=\textwidth]{graphs/n_values_mixedllm2.png}
%         \caption{Lower Intelligence}
%        \label{fig:results:n_values_mixedllm_weak}
%     \end{subfigure}
%     \caption{Frequency of strategic levels for each agent across periods within each set-up.}
%     \label{fig:results:n_values_mixedllm}
%\end{figure}
%\noindent
\vspace{-\baselineskip}

\textit{Payoffs:}
The maximum possible payoffs that can be achieved in the mixed environment is either comparable or could be higher in mixed environments than pure environments.
Since this is a competitive game, higher gain for some also means higher losses for some, thus the variability in payoff outcomes is also larger.
%In set-up 1, even for homogeneous agents, there are some variations in choices, some of them obtained higher payoffs because of a head-start in guessing a slightly lower number at the beginning.
%In set-up 5, the low types do not have much variability and typically behave randomly and in unison, thus they split the prize in each round.
%In view of these, there is greater upside benefit and downside risk in the pure high intelligent environment as compared to the pure low intelligent environment.
While low types usually obtain positive payoffs at the beginning of the game for choosing $50$, which is closer to $\frac{2}{3}$ of the average, this head-start is soon eroded if there exist any high type in the group, who learn to react to this information rapidly, therefore, low types are less likely to win across periods.
The degree of heterogeneity also matters.
High types could obtain higher average payoffs at the expense of low types when low types $\geq 50\%$, and low types are better off if there are less high types.
(Appendix~\ref{detailed_set_up}\label{convergence_llm_return})
%In set-up 3, the higher payoffs achieved by the high types come at the expense of low type agents, who earns $0$ payoffs majority of the time.
%As long as there are intelligent agent(s) in the pool, low intelligent agent(s) can usually earn positive payoff in the first period, but they tend to earn $0$ payoffs thereafter, except in set-up 4, one low type LLM-based agent is able to tie with the high type after period 1, and receive positive payoffs in all 5 periods.

\textit{Application.}
The simulation results could assist in informing policies.
An obvious example is the streaming system in schools, where students are allocated into different classes based on their grades to facilitate better learning. (\cite{ireson1999raising}, \cite{liem2013big})
%Singapore for instance, practice nationwide within-school ability grouping. (\cite{liem2013big})
%There are also extensive literature in this area that explore the impact of such system on students' perception of learning experiences, variations in teachers' expectation, etc. (\cite{joyce2010student, johnston2023students})
%Unlike \cite{joyce2010student, johnston2023students}, my simulation results obstruct from any peer effects or differences in allocation of resources and teachers' attention, 
%Focusing learning rates given variation in class composition.
Suppose students are classified into high and low types in terms of ability, my findings provide an argument for a mixed learning environment, where the low types would learn faster when integrated into a class with larger proportion of higher ability peers;
%, possibly due to revision in beliefs about opponents' actions; 
even for high types, their learning rate could be slightly improved.
%potentially due to the prospect of attaining higher rewards when competing against peers of comparatively lower ability.
%Promoting faster learning could be a practical objective in education.
%\cite{proto2022intelligence} found in repeated Prisoner’s Dilemma games played among human subjects that there are overall higher cooperation rates in the integrated treatment groups, where subjects of different IQ levels interact with one another, as compared to separated treatment groups that have pure high IQ or low IQ subjects.
%Drawing relation to their paper, my results show that in competitive setting, there is also better outcome for integrated treatment groups in terms of higher convergence rates, indicative of more learning taking place, which could be a practical objective in the education context.
Furthermore, the results also makes a case for the usefulness of a variety of LLMs, including weaker models.
Even though they do not learn when competing with each other, they could when placed in presence of stronger LLMs.
Stronger LLMs could also benefit from playing against small proportion of weaker LLMs as shown by higher learning rates, or better average payoffs when playing against higher proportion of weaker LLMs.
%Therefore, playing against weaker opponents could in fact be helpful in improving the performance of the stronger models in strategic situations, 
These set forth the value of continual investment in LLMs of differing strength.
%depending on the objectives of the users, their choices of algorithms to adopt could differ based on the results of strategic interactions described previously.
%If users are looking for short-term positive payoffs, using weaker algorithms could be more attractive (for instance, ~\palm~ usually choose lower number than ~\gptthree~ in period 1);
%on the other hand, if they seek long-term positive payoffs, using stronger algorithms could be more tempting after accounting for the usual higher cost associated with the stronger models.

\textbf{Reasoning Elicitation.}
While it is recognized that drawing direct relations between LLM-based agents and humans in terms of internal reasoning process may be speculative and overextending parallels, therefore analyzing observed actions take precedence in this paper, but reasoning elicitation may serve as an avenue to gain some potential idea of agents' rationale for making certain choices and how they might learn.
In all set-ups, LLM-based agents were prompted at the beginning of period 1 to state their understanding of the game, and for each subsequent periods, they are asked to reinstate the goals.
This step is essential to mitigate the potential of them not comprehending the game.
In which case, LLM-based agents are able to correctly recite the game rules.
The agents were also asked to give a statement of reasoning in support of their choices.
In period 1, both high and low types make choices based on their belief of a popular number, which is often the mean of the range.
In subsequent periods, I found that low types appear to learn by either adjusting the reference point, and make selection that still comply with a strategic level of $0$, or via imitation by following the winner's past choice.
They may also not learn at all, and continue to select a number that they believe to be the popular choice.
As for the high types, they can learn by (1) anchoring their guesses to two-thirds of the past period's average; (2) imitating winner's strategy; (3) adjusting based on past period payoffs; and also (4) pattern recognition.
Agents may place different reliance on distinct pieces of historical information when making their choices, and multiple types of learning could come into play.
This diversity in learning mechanisms could lead to higher speed of changes in average choices, and in turn translate into higher strategic level.
(Appendix~\ref{reasoning}\label{reasoningreturn})

\section{Limitations and Extensions}
\label{sec:extensions}

Much like experiments with human subjects, LLM-based agents could also be sensitive to variations in game design, feedback, and instructions.
This work only explored a small number of set-ups and for a particular competitive game, which can be a limitation in scope, but it serves the main purpose of pitching for the potential of LLMs as a valuable tool for social sciences research, and beauty contests being a game of substantial impact in economics research provides an excellent foundation for this line of work.
Some of the possible extensions would be to include:

\textbf{Variations in Game Design and Feedback.}
While I focused on $p=\frac{2}{3}$, $p$ can be varied to $\frac{1}{2}$ or $\frac{4}{3}$ to replicate human subject experiments, in which case, equilibrium multiplicity could arise, allowing for analysis on equilibrium selection. (\cite{nagel1995unraveling})
In addition, the same set-ups can be implemented but with variations in which piece(s) of historical information to reveal.

\textbf{Objectives.}
Human are sensitive to problem framing and phrasing of survey questions, similarly, LLMs' decisions could be influenced by the formatting of prompts as well. (\cite{tversky1981framing}, \cite{kalton1982effect}, \cite{sclar2023quantifying})
This work explores how agents behave when the objectives are set to be winning the game and followed by maximizing their payoffs, but in most economic models, the primary focus is usually on maximizing utilities and then winning.
In this competitive game, the winning strategy is also one that gives the best payoff, thus changing the sequence of objectives is unlikely to result in drastic differences in game outcomes, but could serve as a sanity check.

\textbf{Prompt Language.}
In \cite{guo2024economics}, the prompt language was changed to Mandarin Chinese in the multi-LLM-based agents setting.
It was found that ~\palm~ is unable to complete the games, indicating potential difficulty in comprehending the instructions when they are given in another language.
As for ~\gptthree~, it can complete the game in Chinese setting but the choices are more clustered.
The variance in strategies observed as compared to the English setting may reflect differences in strategic behaviours among different language users that the models are trained on, or it could stem from a significantly smaller availability of human-generated data in another language.
%%which is an area that can be improved on to better represent the population.
While current work focused on English setting, future work could involve replicating the set-ups in other prompt languages to model heterogeneous population in another dimension.

\textbf{Human-machine Interactions.}
Previously, experimental designs involving computers usually comprise of pre-defined algorithms, and humans were found to display higher degree of strategic reasoning when competing against fellow human opponents as opposed to computer algorithms. (\cite{coricelli2009neural})
Human vs. LLMs could offer a fresh form of human-machine interactions as LLM-based agents could respond dynamically and switch their strategies given historical information, thereby contributing to greater strategic uncertainty and complexity.
Given LLMs display some degree of learning abilities, they could also be learning from playing with human subjects, making the interactions more intriguing to explore.

\textbf{Future Validity.}
Another important question would be the future validity of the results proposed by this paper.
In this paper, the measures of strategic levels are robust to changing game parameter, such as the upper bound of the choice range, which could serve as a form of sensitivity test and make the results more replicable under the same conditions.
%Even though specified versions of LLMs are used for simulation of various set-ups, future versions of LLMs can be easily incorporated under the same framework.
Apart from this, there has been growing interest in exploring if prompting LLMs with questions could make them more strategically sophisticated in the future, and therefore the results cannot be replicated.
This work shows that within a given session, models converge towards NE choice if they gain exposure to past play information, which is indicative of their learning ability over time, offering the possibility of individuals training their own algorithms to better fit their preferences in different contexts and LLMs becoming more sophisticated in the future.
However, since the experiments are conducted with effectively stagnant LLM versions, and information in each round is controlled and has no impact on the back-end databases, this enhance the validity of results, making them replicable under the same set-ups.
If future versions of LLMs incorporate the questions asked by the individuals into their training, then new models could be relatively more sophisticated or on the contrary, less sophisticated due to incorporation of more noisy data.
This could give rise to more serious concerns over whom are the models aligning to, which is an open question for future exploration.

\section{Conclusion}
\label{sec:conclusion}

The contribution of this work is threefold.
Firstly, it serves as part of the literature that seeks to make a case for integration of LLMs as tools for social sciences research.
It then proposed the adoption of LLM-based agents in multi-player competitive games and explore the beauty contests in particular.
Drawing parallels to human subjects, LLM-based agents were evaluated similarly in terms of strategic levels and were found to have levels in between $0$ and $1$, which is slightly lower than human subjects.
Most of them also exhibit learning from historical information, showing convergence to the NE choice at varying rate, demonstrating either revision in ``beliefs" about their opponents, or increasing depth of reasoning.
Similar to human subjects, though strategic levels evolve over time, the increase is minimal.
Further, to better understand strategic interactions in varying environments, I simulated game play between LLM-based agents and fixed-strategy opponents, as well as among LLM-based agents.
%When facing static algorithms, high type LLM-based agents show step-by-step convergence towards $0$, while low types either converge straightaway or stick to the period 1 choice.
%When facing fellow LLM-based agents, high types converge to $0$ more rapidly than low types.
By varying the proportion of agent types in each group,
%Under this interpretation, simulated agents converge fast to $0$ when they know they are playing against static algorithms since there is less strategic uncertainty, and they converge slower when playing against fellow simulated agents.
I found LLM-based agents converge slower to $0$ as the proportion of fixed-strategy agents decreases, demonstrating the impact of increased strategic uncertainty.
Agents were also found to learn faster when placed in mixed environments with players of different strategic levels than environments that comprises of a single type.
This postulates the potential for stimulating faster learning, particularly among less intelligent agents, by introducing heterogeneity into the groups.
Last but not least, this work offers some insights into how different algorithms could behave when interacting with one another, showing potential outcomes if algorithms were to act as proxies for humans and applied to competitive situations.

There are many possible extensions and great potentials for LLMs to be employed as toolkits for social sciences research in interpreting and deciphering human behaviour, which remain a relatively new subject area.
The reverse is true as well, theories and experimental results from decades of learning about human decision-making can be similarly used to better understand machine behaviours and improve their performance.

\clearpage
\bibliography{References}

\begin{thebibliography}{}

\bibitem[Aher et~al., 2023]{aher2023using}
Aher, G.~V., Arriaga, R.~I., and Kalai, A.~T. (2023).
\newblock Using large language models to simulate multiple humans and replicate human subject studies.
\newblock In {\em International Conference on Machine Learning}, pages 337--371. PMLR.

\bibitem[Akata et~al., 2023]{akata2023playing}
Akata, E., Schulz, L., Coda-Forno, J., Oh, S.~J., Bethge, M., and Schulz, E. (2023).
\newblock Playing repeated games with large language models.
\newblock {\em arXiv preprint arXiv:2305.16867}.

\bibitem[Argyle et~al., 2023]{argyle2023out}
Argyle, L.~P., Busby, E.~C., Fulda, N., Gubler, J.~R., Rytting, C., and Wingate, D. (2023).
\newblock Out of one, many: Using language models to simulate human samples.
\newblock {\em Political Analysis}, 31(3):337--351.

\bibitem[Bauer et~al., 2023]{bauer2023decoding}
Bauer, K., Liebich, L., Hinz, O., and Kosfeld, M. (2023).
\newblock Decoding gpt’s hidden ‘rationality’of cooperation.

\bibitem[Bosch-Domenech et~al., 2002]{bosch2002one}
Bosch-Domenech, A., Montalvo, J.~G., Nagel, R., and Satorra, A. (2002).
\newblock One, two,(three), infinity,…: Newspaper and lab beauty-contest experiments.
\newblock {\em American Economic Review}, 92(5):1687--1701.

\bibitem[Brown and MacKay, 2023]{brown2023competition}
Brown, Z.~Y. and MacKay, A. (2023).
\newblock Competition in pricing algorithms.
\newblock {\em American Economic Journal: Microeconomics}, 15(2):109--156.

\bibitem[Camerer et~al., 2004]{camerer2004cognitive}
Camerer, C.~F., Ho, T.-H., and Chong, J.-K. (2004).
\newblock A cognitive hierarchy model of games.
\newblock {\em The Quarterly Journal of Economics}, 119(3):861--898.

\bibitem[Chen et~al., 2016]{chen2016empirical}
Chen, L., Mislove, A., and Wilson, C. (2016).
\newblock An empirical analysis of algorithmic pricing on amazon marketplace.
\newblock In {\em Proceedings of the 25th international conference on World Wide Web}, pages 1339--1349.

\bibitem[Chen et~al., 2023]{chen2023emergence}
Chen, Y., Liu, T.~X., Shan, Y., and Zhong, S. (2023).
\newblock The emergence of economic rationality of gpt.
\newblock {\em Proceedings of the National Academy of Sciences}, 120(51):e2316205120.

\bibitem[Coricelli and Nagel, 2009]{coricelli2009neural}
Coricelli, G. and Nagel, R. (2009).
\newblock Neural correlates of depth of strategic reasoning in medial prefrontal cortex.
\newblock {\em Proceedings of the National Academy of Sciences}, 106(23):9163--9168.

\bibitem[Costa-Gomes and Weizs{\"a}cker, 2008]{costa2008stated}
Costa-Gomes, M.~A. and Weizs{\"a}cker, G. (2008).
\newblock Stated beliefs and play in normal-form games.
\newblock {\em The Review of Economic Studies}, 75(3):729--762.

\bibitem[Devetag et~al., 2016]{devetag2016eye}
Devetag, G., Di~Guida, S., and Polonio, L. (2016).
\newblock An eye-tracking study of feature-based choice in one-shot games.
\newblock {\em Experimental Economics}, 19:177--201.

\bibitem[Dillion et~al., 2023]{dillion2023can}
Dillion, D., Tandon, N., Gu, Y., and Gray, K. (2023).
\newblock Can ai language models replace human participants?
\newblock {\em Trends in Cognitive Sciences}.

\bibitem[Fan et~al., 2023]{fan2023can}
Fan, C., Chen, J., Jin, Y., and He, H. (2023).
\newblock Can large language models serve as rational players in game theory? a systematic analysis.
\newblock {\em arXiv preprint arXiv:2312.05488}.

\bibitem[Goldberg, 1991]{goldberg1991every}
Goldberg, D. (1991).
\newblock What every computer scientist should know about floating-point arithmetic.
\newblock {\em ACM computing surveys (CSUR)}, 23(1):5--48.

\bibitem[Guo, 2023]{guo2023gpt}
Guo, F. (2023).
\newblock Gpt in game theory experiments.
\newblock {\em arXiv:2305.05516}.

\bibitem[Guo et~al., 2024]{guo2024economics}
Guo, S., Bu, H., Wang, H., Ren, Y., Sui, D., Shang, Y., and Lu, S. (2024).
\newblock Economics arena for large language models.
\newblock {\em arXiv preprint arXiv:2401.01735}.

\bibitem[Horton, 2023]{horton2023large}
Horton, J.~J. (2023).
\newblock Large language models as simulated economic agents: What can we learn from homo silicus?
\newblock Technical report, National Bureau of Economic Research.

\bibitem[HuggingFace, 2022]{huggingface}
HuggingFace (2022).
\newblock Illustrating reinforcement learning from human feedback (rlhf).

\bibitem[Huijzer and Hill, 2023]{huijzer_hill_2023}
Huijzer, R. and Hill, Y. (2023).
\newblock Large language models show human behavior.

\bibitem[IBM, 2024]{ibm}
IBM (2024).
\newblock Tokens and tokenization.
\newblock Accessed: 2024-04-10.

\bibitem[Ireson and Hallam, 1999]{ireson1999raising}
Ireson, J. and Hallam, S. (1999).
\newblock Raising standards: Is ability grouping the answer?
\newblock {\em Oxford review of education}, 25(3):343--358.

\bibitem[Kalton and Schuman, 1982]{kalton1982effect}
Kalton, G. and Schuman, H. (1982).
\newblock The effect of the question on survey responses: A review.
\newblock {\em Journal of the Royal Statistical Society Series A: Statistics in Society}, 145(1):42--57.

\bibitem[Keynes, 1936]{keynes1936general}
Keynes, J.~M. (1936).
\newblock The general theory of interest, employment and money.

\bibitem[Kosinski, 2023]{kosinski2023theory}
Kosinski, M. (2023).
\newblock Theory of mind may have spontaneously emerged in large language models.
\newblock {\em arXiv preprint arXiv:2302.02083}.

\bibitem[Liem et~al., 2013]{liem2013big}
Liem, G. A.~D., Marsh, H.~W., Martin, A.~J., McInerney, D.~M., and Yeung, A.~S. (2013).
\newblock The big-fish-little-pond effect and a national policy of within-school ability streaming: Alternative frames of reference.
\newblock {\em American Educational Research Journal}, 50(2):326--370.

\bibitem[Mauersberger and Nagel, 2018]{mauersberger2018levels}
Mauersberger, F. and Nagel, R. (2018).
\newblock Levels of reasoning in keynesian beauty contests: a generative framework.
\newblock In {\em Handbook of computational economics}, volume~4, pages 541--634. Elsevier.

\bibitem[Mei et~al., 2024]{mei2024turing}
Mei, Q., Xie, Y., Yuan, W., and Jackson, M.~O. (2024).
\newblock A turing test of whether ai chatbots are behaviorally similar to humans.
\newblock {\em Proceedings of the National Academy of Sciences}, 121(9):e2313925121.

\bibitem[Nagel, 1995]{nagel1995unraveling}
Nagel, R. (1995).
\newblock Unraveling in guessing games: An experimental study.
\newblock {\em The American economic review}, 85(5):1313--1326.

\bibitem[Nagel et~al., 2017]{nagel2017inspired}
Nagel, R., B{\"u}hren, C., and Frank, B. (2017).
\newblock Inspired and inspiring: Herv{\'e} moulin and the discovery of the beauty contest game.
\newblock {\em Mathematical Social Sciences}, 90:191--207.

\bibitem[OpenAI, 2024]{openai}
OpenAI (2024).
\newblock How chatgpt and our language models are developed.

\bibitem[Ouyang et~al., 2022]{ouyang2022training}
Ouyang, L., Wu, J., Jiang, X., Almeida, D., Wainwright, C., Mishkin, P., Zhang, C., Agarwal, S., Slama, K., Ray, A., et~al. (2022).
\newblock Training language models to follow instructions with human feedback.
\newblock {\em Advances in Neural Information Processing Systems}, 35:27730--27744.

\bibitem[Phelps and Russell, 2023]{phelps2023investigating}
Phelps, S. and Russell, Y.~I. (2023).
\newblock Investigating emergent goal-like behaviour in large language models using experimental economics.
\newblock {\em arXiv preprint arXiv:2305.07970}.

\bibitem[Sclar et~al., 2023]{sclar2023quantifying}
Sclar, M., Choi, Y., Tsvetkov, Y., and Suhr, A. (2023).
\newblock Quantifying language models' sensitivity to spurious features in prompt design or: How i learned to start worrying about prompt formatting.
\newblock {\em arXiv preprint arXiv:2310.11324}.

\bibitem[Trality, 2024]{trality}
Trality (2024).
\newblock Crypto trading bots: The ultimate beginner's guide.

\bibitem[Tversky and Kahneman, 1981]{tversky1981framing}
Tversky, A. and Kahneman, D. (1981).
\newblock The framing of decisions and the psychology of choice.
\newblock {\em science}, 211(4481):453--458.

\bibitem[Webb et~al., 2023]{webb2023emergent}
Webb, T., Holyoak, K.~J., and Lu, H. (2023).
\newblock Emergent analogical reasoning in large language models.
\newblock {\em Nature Human Behaviour}, 7(9):1526--1541.

\end{thebibliography}
\bibliographystyle{apalike}
%\printbibliography

\newpage

\appendix
\section{Appendix}

\subsection{Original Exemplary Prompt}
\label{originalprompt}

\textbf{One-shot games with multi-LLM-based agents.}
\definecolor{mycolor}{RGB}{255, 223, 186}
\begin{tcolorbox}[colback=mycolor, colframe=mycolor, sharp corners, boxrule=0pt]
\textbf{[System Instruction.]} I want you to act as a clever game player and perform as rationally as possible.

\vspace{12pt}

\textbf{[User Instruction.]}
You are playing a game and the rules are as follows:
    \begin{itemize}
        \item it consists of {\color{blue}{\{number of players\}}} players, including you.
        \item this is a one-round game.
        \item you need to choose a real number between 0 and {\color{blue}{\{upper bound\}}}, 0 and {\color{blue}{\{upper bound\}}} inclusive.
        \item everyone does not know how others would make choices beforehand.
        \item the winner is the person whose number is the closest to the two thirds of the average of all chosen numbers.
        \item the winner gets a fixed prize, in case of a tie the prize is split amongst those who tie.
        \item your goal is primarily to maximize the possibility of getting the prize and secondly to maximize the your prize.
    \end{itemize}
    %(Return to \ref{return_historical_prompt})
\end{tcolorbox}
\noindent
\textbf{Subseqent Prompt for Historical Information.}\label{historical_prompt}
\begin{tcolorbox}[colback=mycolor, colframe=mycolor, sharp corners, boxrule=0pt]
\textbf{[User Instruction.]}
\begin{itemize}
    \item The game of the same config has been hold for {\color{blue}{\{number of runs\}}} run(s), and the historical choices of everyone are shown below (your id is {\color{blue}{\{ID of the agent\}}}:
    {\color{blue}{\{historical information including (1) period index, (2) choices made by all agents, (3) average of the choices; (4) 2/3 of the average; (5) winner id.\}}}
    \item Everyone can optimize his/her answer with the history to play in a new run in order to achieve goals.
\end{itemize}
    (Return to \ref{return_historical_prompt})
    (Return to \ref{return_historical_prompt_llm})
\end{tcolorbox}

\newpage

\subsection{New Exemplary Prompt}
\label{newprompt}

\textbf{Opponents Playing Fixed Strategy of $0$}\label{newprompt_fixedstrategy_NE}
\begin{tcolorbox}[colback=mycolor, colframe=mycolor, sharp corners, boxrule=0pt]
\textbf{[System Instruction.]} I want you to act as a clever game player and perform as rationally as possible.

\vspace{12pt}

\textbf{[User Instruction.]}
You are playing a game and the rules are as follows:
    \begin{itemize}
        \item it consists of {\color{blue}{\{number of players\}}} players, including you.
        \item this is a one-round game.
        \item you need to choose a real number between 0 and {\color{blue}{\{upper bound\}}}, 0 and {\color{blue}{\{upper bound\}}} inclusive.
        \item everyone does not know how others would make choices beforehand.
        \item the winner is the person whose number is the closest to the two thirds of the average of all chosen numbers.
        \item the winner gets a fixed prize, in case of a tie the prize is split amongst those who tie.
        \item your goal is primarily to maximize the possibility of getting the prize and secondly to maximize the your prize.
        \item \color{red}{some of your opponents will be playing a fixed strategy of $0$ and all others are behaving as rationally as possible.}
    \end{itemize}

\textit{Follow-up for each period.}

Please just strictly output a JSON string, which has following keys:
\begin{itemize}
    \item understanding: str, your brief understanding of the game
    \item popular answer: float, the number which you think other players are most likely to choose
    \item answer: float, the number which you would like to choose
    \item reason: str, the brief reason why you give the popular answer and the answer that way
\end{itemize}

\textit{Subsequent Prompt (after period 1).}
\begin{itemize}
    \item The game of the same config has been hold for {\color{blue}{\{number of runs\}}} run(s), and the historical choices of everyone are shown below (your id is {\color{blue}{\{ID of the agent\}}}:
    {\color{blue}{\{historical information including (1) period index, (2) choices made by all agents, (3) average of the choices; (4) 2/3 of the average; (5) winner id.\}}}
    \item Everyone can optimize his/her answer with the history to play in a new run in order to achieve goals.
\end{itemize}

    (Return to \ref{return_newprompt_fixedstrategy_NE})
\end{tcolorbox}
%\\ \ \\
%\textbf{One-shot games with Hard-coded Fixed Strategy Opponents that Plays the NE Strategy.}
%\begin{tcolorbox}[colback=mycolor, colframe=mycolor, sharp corners, boxrule=0pt]
%\textbf{[System Instruction.]} I want you to act as a clever game player and perform as rationally as possible.
%
%\vspace{12pt}
%
%\textbf{[User Instruction.]}
%You are playing a game and the rules are as follows:
%    \begin{itemize}
%        \item it consists of {\color{blue}{\{number of players\}}} players, including you.
%        \item this is a one-round game.
%        \item you need to choose a real number between 0 and {\color{blue}{\{upper bound\}}}, 0 and {\color{blue}{\{upper bound\}}} inclusive.
%        \item everyone does not know how others would make choices beforehand.
%        \item the winner is the person whose number is the closest to the two thirds of the average of all chosen numbers.
%        \item the winner gets a fixed prize, in case of a tie the prize is split amongst those who tie.
%        \item your goal is primarily to maximize the possibility of getting the prize and secondly to maximize the your prize.
%        \item \color{red}{you can assume that other are all perfectly rational players.}
%    \end{itemize}
%    (Return to \ref{returnoriginalprompt1})
%\end{tcolorbox}

\newpage

\subsection{Additional Details}

\subsubsection{Choice Variability Given the Same Upper-bound}
\label{choicevariability}
\begin{wrapfigure}{r}{0.5\textwidth}
    \centering
    \includegraphics[width=5.5cm, height=4.2cm]{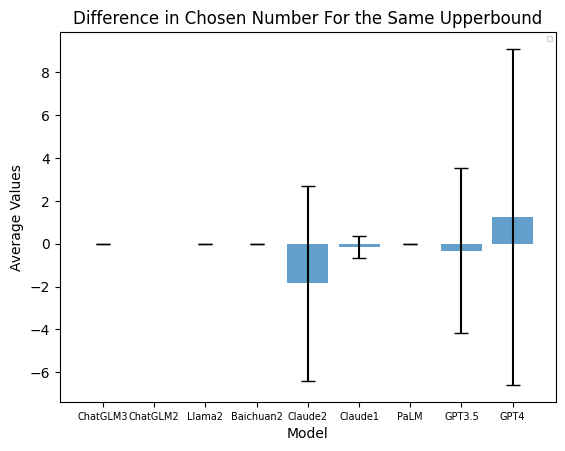}
    %{graphs/same_upperbound.png}
    \caption{Variability in chosen number given the same upper-bound.
    %Slightly larger variability for stronger models.
    }
    \label{fig:results:same_upperbound_diff_sessions}
\end{wrapfigure}
%This exercise is important to determine if, like human players, there could be variability in choices.
%Within the 150 sessions, by looking at the cases where upper-bound is the same, the difference between the choices made are computed as in 
Figure~\ref{fig:results:same_upperbound_diff_sessions} shows that within the 150 sessions, for the sessions that have the same randomly generated upper-bound, $\bar{c}$, the same LLM-based agent could choose slightly different numbers.
For instance, ~\claudetwo~, ~\gptthree~ and ~\gptfour~ displayed more variability in choices as compared to other models.
This results is indicative that, like human players, there could be variability in choices for LLM-based agents.
Since choices might not be static even when the instructions is exactly the same, the determination of average choices and the corresponding strategic levels based on both identical and different upper-bounds would lead to a more consistent and robust measure.

Return to Section~\ref{choicevariabilityreturn}.

\subsubsection{Choices and Strategic Levels in Repeated Games}
%%In the repeated setting, it is possible to track for changes over time.
\begin{figure}[H]
    \centering
    \begin{subfigure}[b]{0.39\textwidth}
         \centering
         \includegraphics[width=\textwidth]{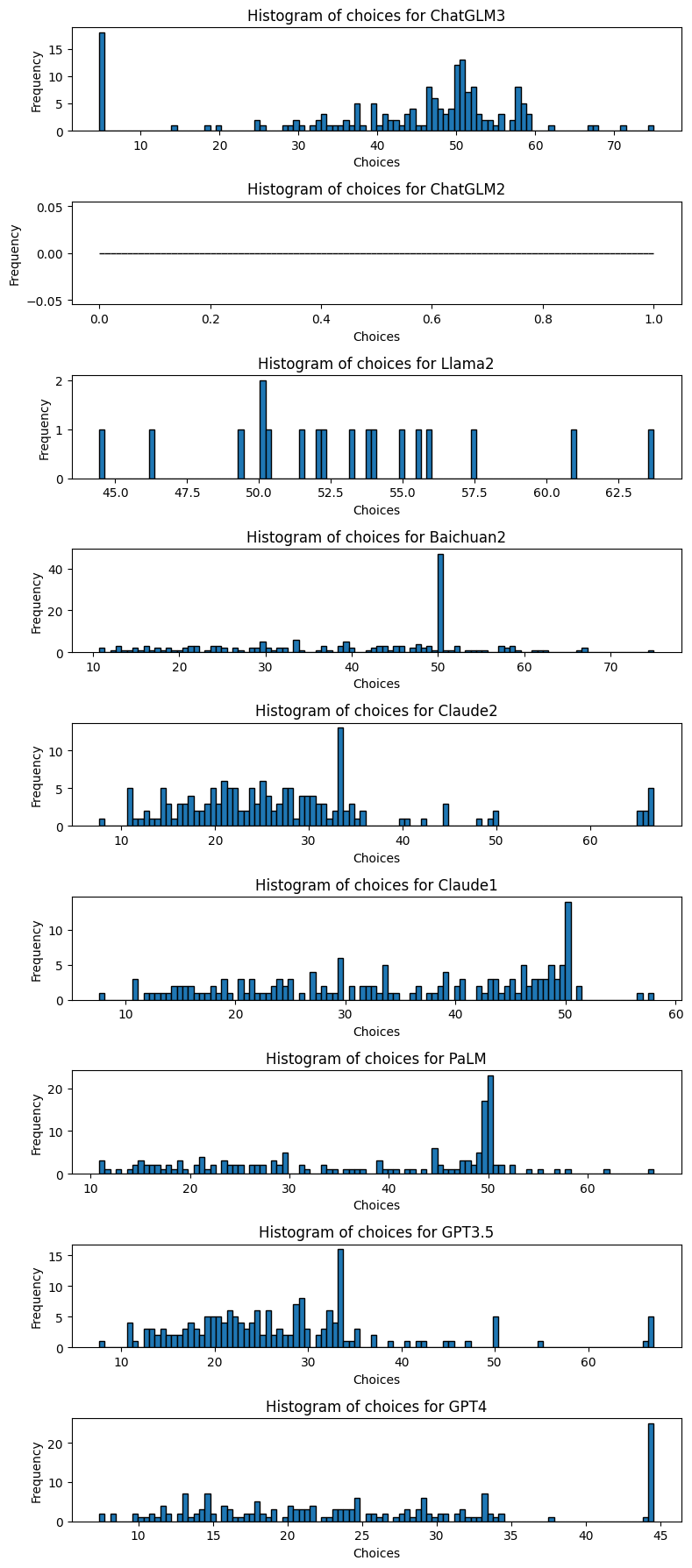}
         \caption{Choices.}
        \label{fig:results:freq_choices}
     \end{subfigure}
     ~
     \begin{subfigure}[b]{0.39\textwidth}
         \centering
    \includegraphics[width=\textwidth]{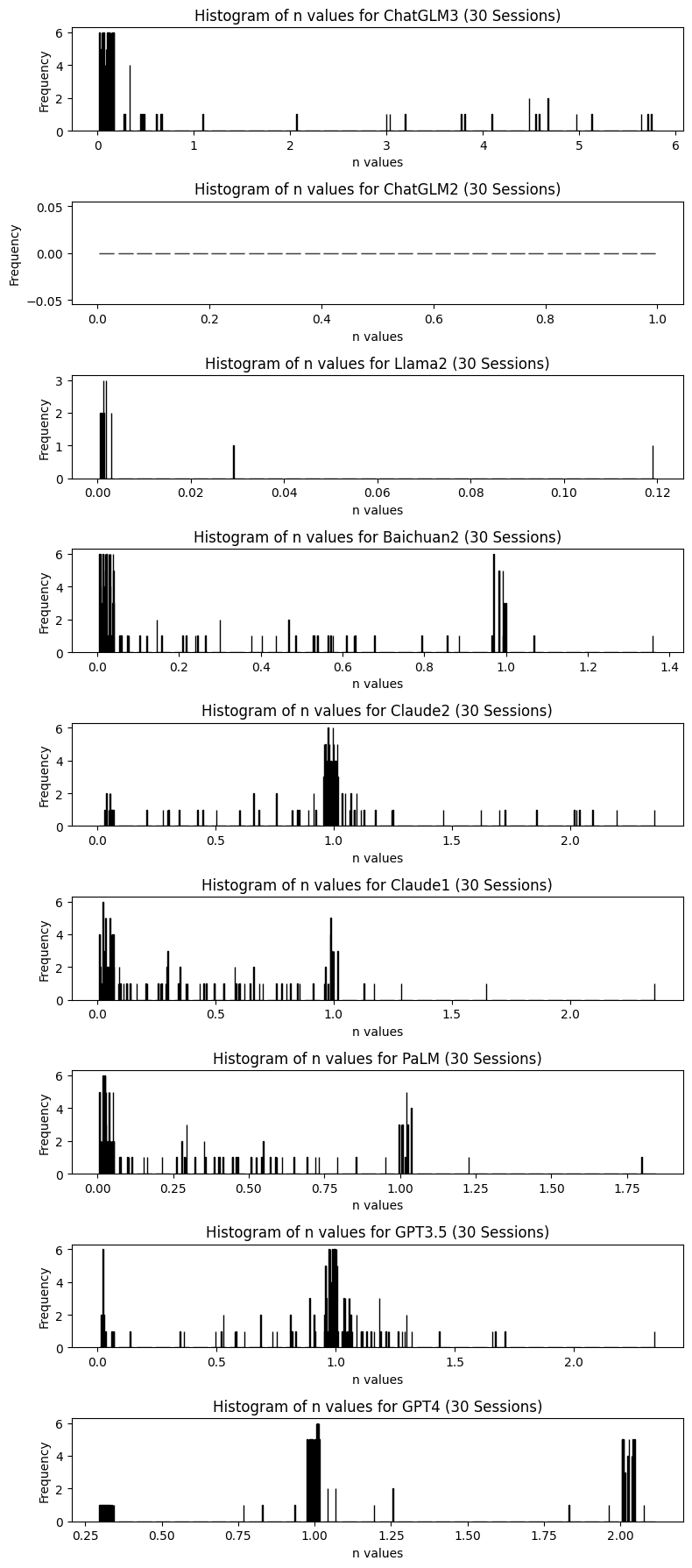}
    \caption{Strategic levels.}
    \label{fig:results:freq_n}
    \end{subfigure}
    \caption{Choices and strategic levels over 6 periods across 30 sessions.}
     \label{fig:results:hist_choices_n}
\end{figure}
\begin{wrapfigure}{r}{0.5\textwidth}
    \centering
    \includegraphics[width=6.5cm, height=4cm]{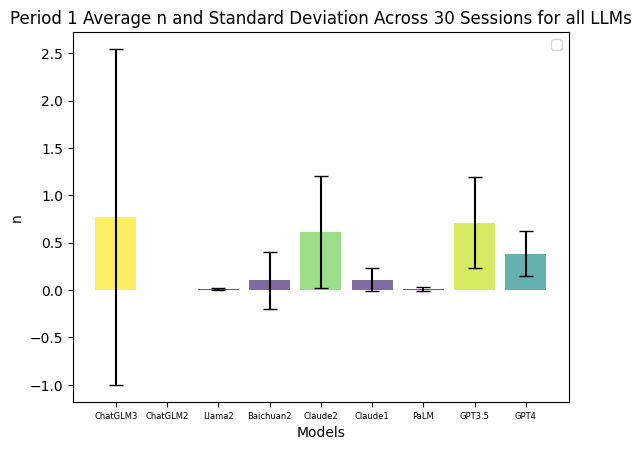}
    \caption{Period 1 strategic level.}
    \label{fig:results:period1_average_n}
\end{wrapfigure}
Figure~\ref{fig:results:hist_choices_n} illustrates the frequency of choices and the corresponding strategic levels computed over $6$ periods and across $30$ sessions for the repeated setting.
The results are similar to the one-shot game, where most of the LLM-based agents choose $50$ (normalized) with high frequency.
There is greater dispersion in number selected as compared to the one-shot games, which is an indication that agents do change their actions over time and there is more selection below $50$.
Following \cite{nagel1995unraveling}'s method of computing for strategic levels in repeated setting, where 
%%%the first period reference point is fixed at $\frac{\bar{c}}{2}$, and in subsequent periods, the 
reference points in period 2 onwards are re-calibrated to be the mean choice of the past period.
In experiments with human subjects, 
there are no support for increasing depth of reasoning as
%%most subjects remain in the bounds of iteration step $0$ and $3$, and 
they rarely go over iteration step $2$.
With LLM-based agents, 
%%based on the frequency of strategic levels computed across periods in Figure~\ref{fig:results:freq_n}, 
the results are similar.
%%they do not go over iteration step $2$, and 
Most of them display spike at $0$ and $1$, with ~\gptfour~ showing spike at both $1$ and $2$.
While they do stay below iteration step $2$, as compared to their period 1 performance (Figure~\ref{fig:results:period1_average_n}), some of them display minor increase in strategic levels.
In particular, ~\gptfour~ has strategic level much lower than $2$ in period 1, but is able to achieve approximately level-$2$ over time.

Return to Section~\ref{choicevariabilityreturn}.

\subsubsection{Variations in Group Composition}
\textbf{Detailed set-ups.}
\label{detailed_set_up}
\begin{itemize}
    \item 10 agents are playing in each game.
    \item The same group plays for 5 periods, and all history are revealed.
    \item They choose a number between 0 and $\bar{c}$, $\bar{c}$ is fixed to be $100$. The winner is the agent whose number is the closest to $p$ times the average of all chosen numbers, where $p=\frac{2}{3}$ to ensure a unique interior NE solution.
    \item In each period, the winner gets a fixed prize of $\$x$. In case of a tie, the prize is split amongst those who tie. All other players receive $0$.
\end{itemize}
%Return to Section~\ref{detailed_set_up_return}.

\textbf{Expected choice variation across periods when playing against fixed-strategy opponents.}
\label{choice_variations_fixed_strategy}
Denoting $a_t$ to be the action/number guessed in each time period, $N_f$ to be the number of fixed-strategy players and $N_l$ to be the number of LLM-based agents, the selection in the next period:
\begin{equation}
    a_{t+1}=BR(N_f, N_l, a_t)=\frac{2}{3}(\frac{N_f}{10}*0+\frac{N_l}{10}a_t)
\end{equation}
The choice variation over the periods is computed with $\frac{a_{t+1}}{a_{t}}$.
There are three treatment groups for LLM-based agents vs. fixed-strategy opponents, differing in proportion of player types.
For 9/10 fixed-strategy agents, the next period guess is expected to be $0.067$ of the previous number; 
For 5/10 fixed-strategy agents, the guess is expected to be $0.333$ of the previous number; 
For 1/10 fixed-strategy agents, the guess is expected to be $0.6$ of the previous number.
Lowering proportion of fixed-strategy types in the group is hypothesized to induce higher guesses and will slow down the convergence process.

\textbf{Expected choice variation across periods when playing against LLM-based agents.}
\label{choice_variations_llm}
Let the strategy of high type in period $t$ be $a_{Ht}$ and that of low type be $a_{Lt}$, the selection in the next period:
\begin{equation}
    a_{it+1}=BR(B(N_H), B(N_L), a_t)=\frac{2}{3}(\frac{B(N_H)}{10}a_{Ht}+\frac{B(N_L)}{10}a_{Lt}), i \in (H, L)
\end{equation}
where $B(N_H)$ and $B(N_L)$ are agent $i$'s ``beliefs" about the number of high types and low types.
When playing against fixed strategy opponents, it is possible to observe in period 2 who selected $0$, thereby deriving the correct proportion of fixed strategy players within the population.
However, as all agents are LLM-based in this set-up, it could be harder to distinguish the proportion of types within the group based on historical choices in period 2, for instance, even if they chose the same number it does not imply they are of the same type.
Further, the agents were not told explicitly their own type relative to the others, so they have to guess if they fall within $N_H$ or $N_L$.
As a result, the best response of a specific agent would be dependent on its ``beliefs" about the proportion of high and low types.
In the case where beliefs are correct given revealed information, then $B(N_H)=N_H$ and $B(N_L)=N_L$.

Suppose one correctly perceived the proportion of agent types based on revealed historical choices, the variation of number selected over the periods could similarly be computed with $\frac{a_{t+1}}{a_{t}}$.
There are five treatment groups for LLM-based agents vs. LLM-based agents, differing in proportion of player types.
\begin{itemize}
    \item Pure high intelligence environment: It is expected that the next period guess will be 0.667 of the previous number.
    \item Highly intelligent environment: \begin{equation*}
        \frac{a_{Ht+1}}{a_{Ht}}=0.067\frac{a_{Lt}}{a_{Ht}}+0.6, \frac{a_{Lt+1}}{a_{Lt}}=0.6\frac{a_{Ht}}{a_{Lt}}+0.067, \text{for } \frac{a_{Ht}}{a_{Lt}}<1, \frac{a_{Ht+1}}{a_{Ht}}>\frac{a_{Lt+1}}{a_{Lt}}
    \end{equation*}
    \item Mixed intelligent environment:
    \begin{equation*}
       \frac{a_{Ht+1}}{a_{Ht}}=0.333\frac{a_{Lt}}{a_{Ht}}+0.333, \frac{a_{Lt+1}}{a_{Lt}}=0.333\frac{a_{Ht}}{a_{Lt}}+0.333, \text{ for } \frac{a_{Ht}}{a_{Lt}}<1, \frac{a_{Ht+1}}{a_{Ht}}>\frac{a_{Lt+1}}{a_{Lt}}
    \end{equation*}
    \item Less intelligent environment:
    \begin{equation*}
       \frac{a_{Ht+1}}{a_{Ht}}=0.6\frac{a_{Lt}}{a_{Ht}}+0.067, \frac{a_{Lt+1}}{a_{Lt}}=0.067\frac{a_{Ht}}{a_{Lt}}+0.6, \text{ for } \frac{a_{Ht}}{a_{Lt}}<1, \frac{a_{Ht+1}}{a_{Ht}}>\frac{a_{Lt+1}}{a_{Lt}}
    \end{equation*}
    \item Pure low intelligence environment: It is expected that the next period guess will be 0.667 of the previous number.
\end{itemize}
\noindent
For pure environments, the rate of change in choices is expected to be the same for high and low types.
For set-ups $2$ to $4$, if high types chose a smaller number than low types because they go through more iterations of reasoning, and $\frac{a_{Ht}}{a_{Lt}}<1$, then high types are expected to lower their guesses less from time $t$ to $t+1$ as compared to low types.
There could mean slower rate of change for high types than low types.
Otherwise, if high types have strong beliefs that they are playing against opponents who will choose higher numbers while low types believe the other way around, then it is possible for $\frac{a_{Ht}}{a_{Lt}}>1$, then low types are expected to lower their guesses less from time $t$ to $t+1$ as compared to high types, implying faster rate of change in choices for high types than low types.

\textbf{Payoff transition when playing against fixed strategy opponents:}
\begin{figure}[H]
     \centering
     \begin{subfigure}[b]{0.31\textwidth}
         \centering
         \includegraphics[width=\textwidth]{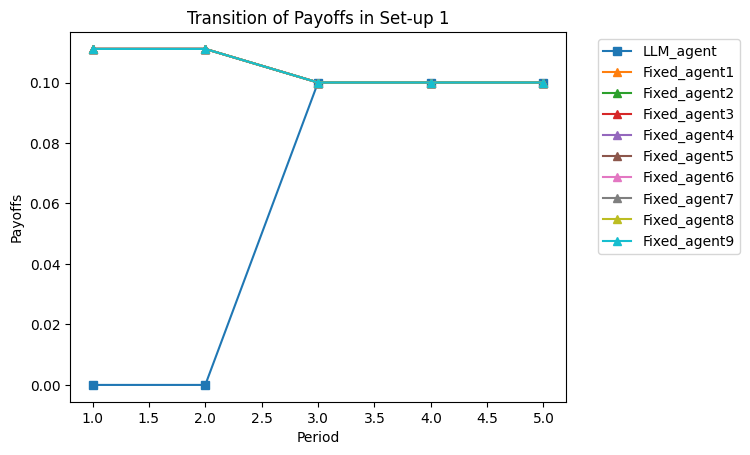}
         \caption{Low}
         \label{fig:payoffs_setup1_static_strong}
     \end{subfigure}
     ~
     \begin{subfigure}[b]{0.31\textwidth}
         \centering
         \includegraphics[width=\textwidth]{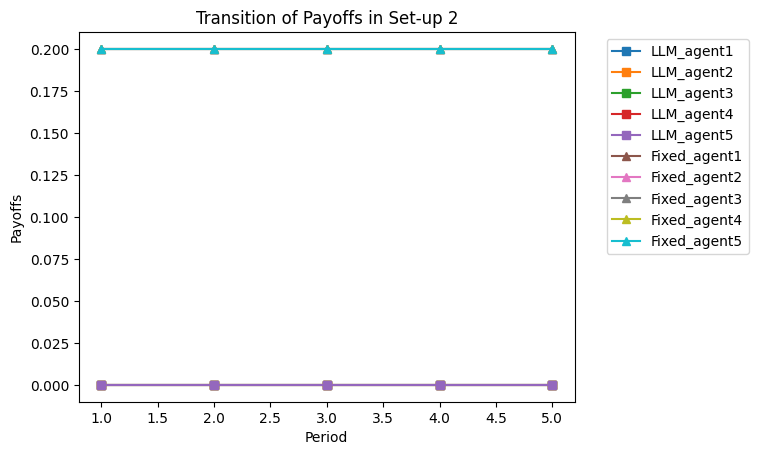}
         \caption{Mixed}
        \label{fig:payoffs_setup2_static_strong}
     \end{subfigure}
     ~
     \begin{subfigure}[b]{0.31\textwidth}
         \centering
         \includegraphics[width=\textwidth]{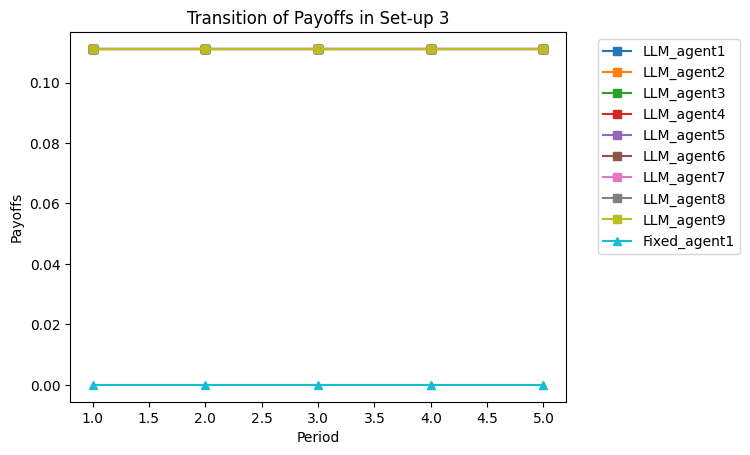}
         \caption{High}
         \label{fig:payoffs_setup3_static_strong}
     \end{subfigure}
     ~
     \\
     \caption{Transition of payoffs for high type LLM-based agent(s) vs. fixed-strategy opponents.}
     \label{fig:payoffs_static_strong}
\end{figure}
\begin{figure}[H]
     \centering
     \begin{subfigure}[b]{0.31\textwidth}
         \centering
         \includegraphics[width=\textwidth]{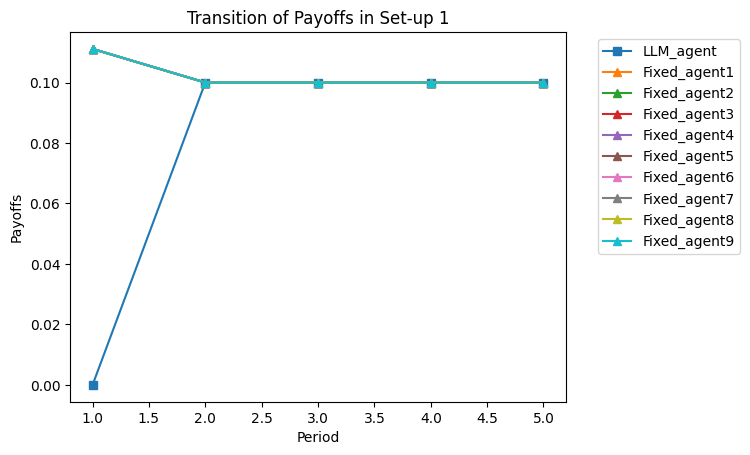}
         \caption{Low}
         \label{fig:payoffs_setup1_static_weak}
     \end{subfigure}
     ~
     \begin{subfigure}[b]{0.31\textwidth}
         \centering
         \includegraphics[width=\textwidth]{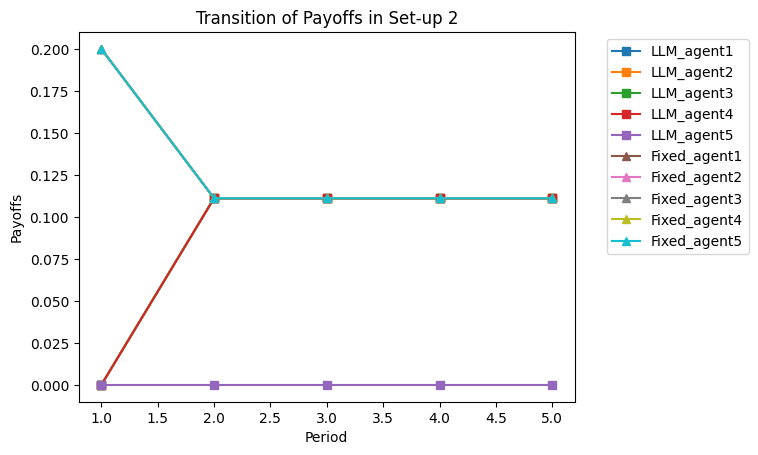}
         \caption{Mixed}
        \label{fig:payoffs_setup2_static_weak}
     \end{subfigure}
     ~
     \begin{subfigure}[b]{0.31\textwidth}
         \centering
         \includegraphics[width=\textwidth]{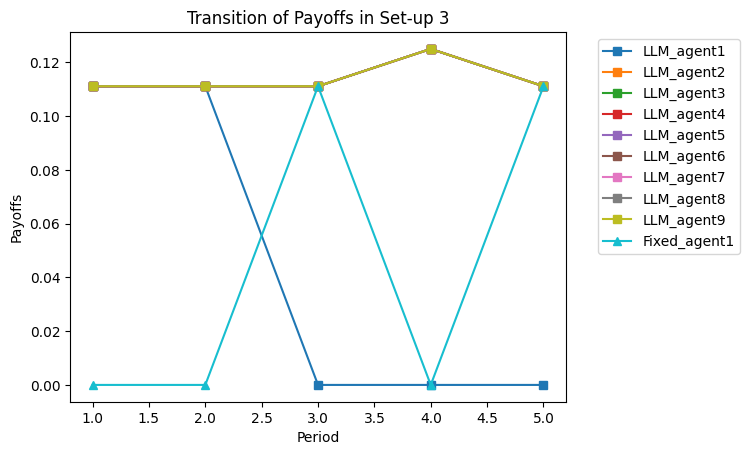}
         \caption{High}
         \label{fig:payoffs_setup3_static_weak}
     \end{subfigure}
     ~
     \\
     \caption{Transition of payoffs for low type LLM-based agent(s) vs. fixed-strategy opponents.}
     \label{fig:payoffs_static_weak}
\end{figure}
%Return to Section~\ref{convergence_fixed_return}.

\textbf{Payoff transition when playing with LLM-based opponents:}
\begin{figure}[H]
     \centering
     \begin{subfigure}[b]{0.31\textwidth}
         \centering
         \includegraphics[width=\textwidth]{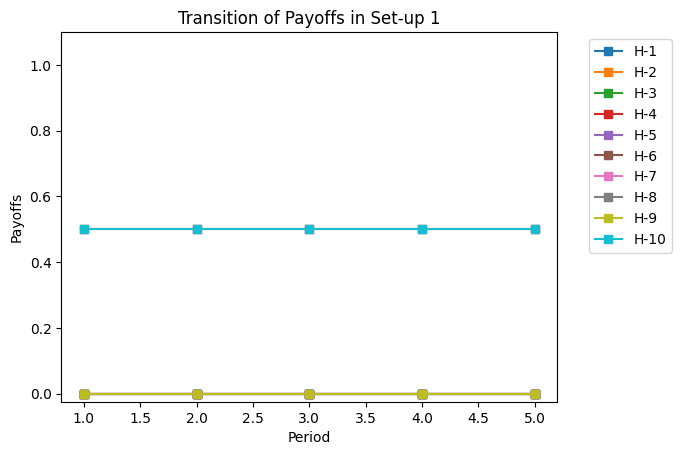}
         \caption{Pure High Intelligent}
         \label{fig:payoffs_setup1_mixedllm}
     \end{subfigure}
     ~
     \begin{subfigure}[b]{0.31\textwidth}
         \centering
         \includegraphics[width=\textwidth]{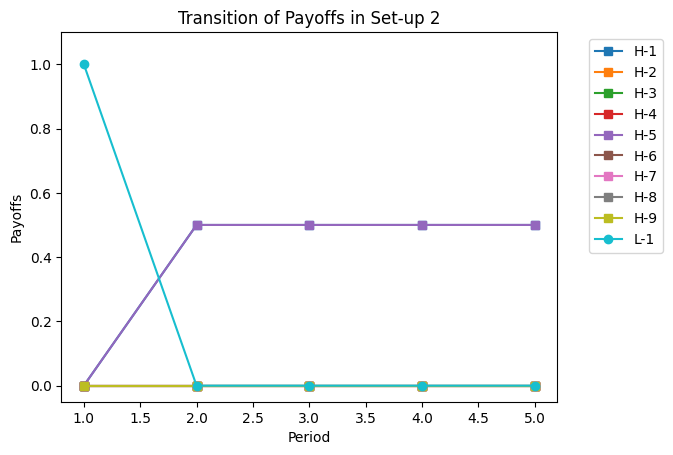}
         \caption{Highly Intelligent}
        \label{fig:payoffs_setup2_mixedllm}
     \end{subfigure}
     ~
     \begin{subfigure}[b]{0.31\textwidth}
         \centering
         \includegraphics[width=\textwidth]{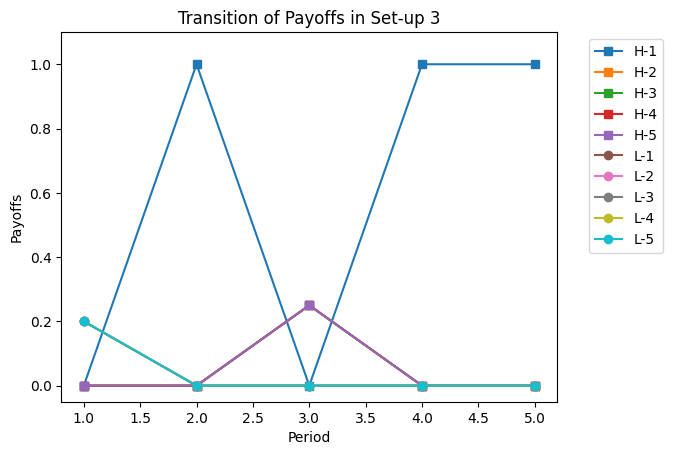}
         \caption{Mixed Intelligent}
         \label{fig:payoffs_setup3_mixedllm}
     \end{subfigure}
     ~
     \begin{subfigure}[b]{0.31\textwidth}
         \centering
         \includegraphics[width=\textwidth]{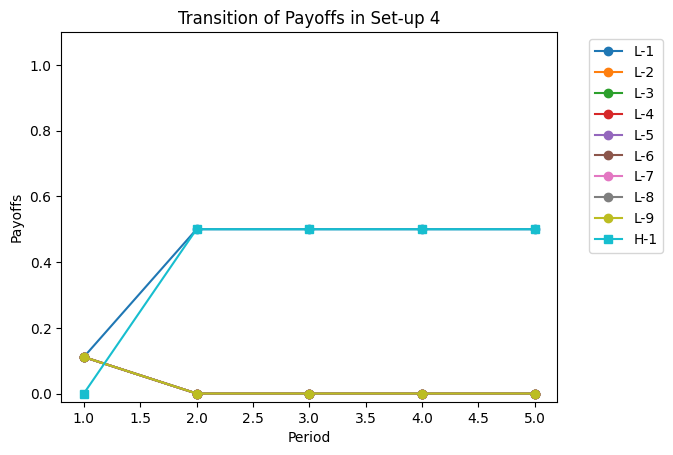}
         \caption{Less Intelligent}
        \label{fig:payoffs_setup4_mixedllm}
     \end{subfigure}
     ~
     \begin{subfigure}[b]{0.31\textwidth}
         \centering
         \includegraphics[width=\textwidth]{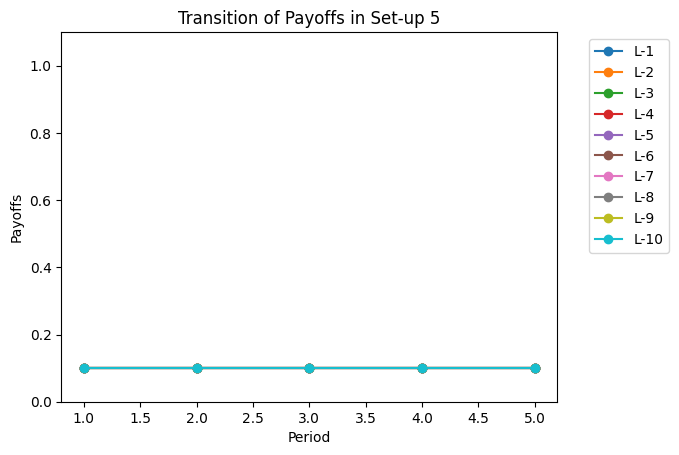}
         \caption{Pure Low Intelligent}
         \label{fig:payoffs_setup5_mixedllm}
     \end{subfigure}
     \\
     \caption{Transition of payoffs given variation in group composition for LLM-based agents playing against each other.}
     \label{fig:payoffs_mixedllm}
\end{figure}

Return to Section~\ref{adaptivelearning}.

\subsubsection{Reasoning Elicitation}\label{reasoning}
It is acknowledged that delve into the reasoning behind LLM-based agents' behaviours could be compelling and beneficial in understanding the choice process better, however, drawing a direct parallel to human subjects in terms of internal reasoning process that actually took place may be speculative.
Therefore, details of the reasoning elicitation results are included as part of the Appendix.
%%In all set-ups, LLM-based agents were given a prompt at the beginning of period 1 to state their understanding of the game, and for each subsequent periods, they are asked to reinstate their goal.
%%This step is essential to mitigate the potential of them not comprehending the game or having incorrect interpretations of the game rules, which could affect how they behave.
%based on erroneous understanding.
%%After this, they are asked to make their choices, followed by giving their line of reasoning.
%%\\ \ \\
For illustration, herein, I explore specifically set-up 2 of LLMs vs. static algorithm and set-up 3 of LLMs vs. LLMs.
The line of reasoning are fairly consistent across different set-ups.
%%so whichever set-up was chosen would not impact the analysis.
%%The main reason for selecting these two set-ups is to focus on an interesting scenario that involve strategic interactions between different types of agents, also they are completely mixed environments with the same number of different types of agents, such that there are no disproportionate influence of one specific type on the average behaviour.

\textit{Partial static environment.}
%%%In the set-up where LLMs are going against static algorithms, LLM-based agents were explicitly told that they are playing with some fixed strategy opponents that select $0$, but the proportion of which is unknown.
In period 1, both $H$ and $L$ agents correctly recite the game rules and objectives they need to follow.
Upon selecting their choices, they were asked about how they arrive at their selection.
Based on the responses, $H$ and $L$ agents indicate that they make their decision based on their perception of the popular number that may be picked, which implies that they are both reacting to beliefs about opponents' choices.
$L$ agents chose $50$, which they believe is the popular answer.
%%to maximize their chances of winning.
$H$ agents chose $66.67$ or $66.6$ because it is close to the upper limit of the range, possibly using the upper-bound as the focal point, or that they are going by iterated elimination of dominated strategies.
Based on period 1 reasoning, 
%%the LLM-based agents show slightly different perception about the behaviour of their opponents.
$L$ agents exhibit strategic naivete and they did not really take into consideration that they are playing with fixed strategy players who will be choosing $0$, they do appear more concern about the average.
Conversely, $H$ agents display more strategic sophistication in reasoning about their opponents' behaviours, and it appears they contemplate more based on opponents' choices and $\frac{2}{3}$ of the average.
%but we can imply from the answers that agents back-boned by different LLMs do possess different belief about their opponents, and while there are more variability in the phrasing of answers by LLM-based agents back-boned by ~\gptthree~, this is as a result of word generation process, the gist of the message is in fact the same.
%%\subsubsection{Subsequent Periods}
%%
%%\textbf{Understanding.} At the beginning of each subsequent periods, the LLM-based agents' understanding of game rules and objectives are verified again for consistency.
%%While the LLM-based agents correctly recite the goal of the game, the high types display more variability in wordings and can sometimes provide a different answer.
%%They could state their goal to be choosing a number closest to two-thirds of the average of all chosen numbers, instead of maximizing their winning probability or to be maximizing their prize.
%%While the two objectives should give rise to the same results and can be perceived as different ways of phrasing the same problem, the linguistic formulation of the former relies less on the specifics of payoffs.
%~\palm~: ``My goal is to maximize the possibility of getting the prize and secondly to maximize the size of my prize."
%\\ \ \\
%~\gptthree~: ``The goal of the game is to choose a number that is closest to two-thirds of the average of all chosen numbers." (22.209333333333333)
%``My goal is still to maximize the possibility of winning the prize and secondarily to maximize my own prize." (22.209)
%%\\ \ \\

In subsequent periods, 
%%In periods after the first, LLM-based agents are able to observe the past choices of their opponents, they can do some computations and adjust their strategies accordingly.
%%Otherwise, they could also obstruct from any calculations, and base their guesses on the information of past average choices or $\frac{2}{3}$ of the past average instead.
%%\\ \ \\
$L$ agents show two possible responses.
The first is to acknowledge the average of all chosen numbers, and that the winner was the one that chose $0.0$.
They would then choose $0.0$, believing it to be the best strategy and the most likely number to be chosen by the other players.
%%thus giving the highest possibility of winning.
The second possible response is to stick to the answer of $50$, believing it to remain the most popular answer, in which case,
%%since it falls in the middle of the range, and choosing it will maximize the chance of winning.
past information does not appear to have any influence on their behaviour.
%\\ \ \\
%~\palm~: ``In the previous run, the average of all chosen numbers was 0.0. This means that the winner was the person who chose 0.0. In this run, I will choose 0.0 as well, because it is the most likely number to be chosen by other players. If I choose 0.0, I will have a 10\% chance of winning, which is the highest chance of winning of any number I could choose."
%OR
%``I think the most popular answer will still be 50, because it is the middle number between 0 and 100. If everyone chooses 50, the average will be 50, and the winner will be the person who chooses 50. However, if I choose 50, I will not be guaranteed to win, because someone else might choose 50 as well. In order to maximize my chances of winning, I should choose a number that is close to 50, but not equal to 50. I could choose 49 or 51, but I think 50 is the best choice, because it is the most likely number to be chosen by other players. If I choose 50, I will have a 10\% chance of winning, which is the highest chance of winning of any number I could choose."
%%\\ \ \\
As for $H$ agents, they acknowledged in their reasoning process that historical information are taken into account, particularly the average and $\frac{2}{3}$ of this average,
%%was found to be $33.314$, and they compute for two-third of this average 
and its choice is adjusted to this value to maximize the probability of winning the game.
%%The adjustment were of similar magnitude among agents of the same type but different entities, the only difference is in terms of the decimal places that one appears to be accounting for.
One discrepancy between different $H$ agents is the number of decimal places they appear to account for when computing past period average. 
For example, after learning period 1 information, some $H$ agents in period 2 adjust to select $22.209333333333333$ and some to $22.209$.
In the set-ups outlined in this paper, agents can choose any number within the range, and computationally, as a result of floating-point precision, the number of choices are finite.
(\cite{goldberg1991every})
The difference in the number selection could be a result of token limitation, which specify the capacity of a model to handle or generate text within a sequence.
(\cite{ibm})
Given the slight variations in responses, the same type of agent could generate numbers of marginally different precision, where they could be rounded off or truncated to fit within the limit.
Even though the difference between the numbers selected are essentially trivial, this property of LLMs may be perceived as a potential rather than a flaw that can have strategic significance and make a difference in payoffs obtained.
%%accounting for different number of decimal places could constitute another potential type of model capability that could have strategic significance and make a difference in payoffs obtained.
However, this is not an issue in this set-up since the fixed-strategy players are the ones winning.

Based on the line of reasoning, $L$ agents either adjust their choices according to the past periods' average or the winners' strategy, which can be an indication of learning through imitation.
It is also possible they show no adjustment and continue to pick $50$, which they perceive to be the popular choice at the start of the game.
%One phrase that kept coming up in their answers is they will form an opinion about the popular choice that will be choose by the other players, and they will choose that popular choice.
%%This could imply that they are following level-0 thinking process, and their unwavering, perceived popular choice is the most important piece of information in determining their choices.
%where they are picking the number that corresponds to the average or their perceived popular choice.
Further, throughout the reasoning process, they did not mention fixed-strategy players, except when prompting their understanding of the game in period 1.
%%It is expected that they learn about the proportion of fixed-strategy opponents after revelation of historical information at the start of period 2, but 
It appears that they do not make use of such information.
On the other hand, $H$ agents adjust their choice according to the information on past periods' average and $\frac{2}{3}$ of that value.
%On the other hand, for LLM-based agents back-boned by ~\gptthree~, they are adjusting according to information on past periods' average and two-thirds of the average.
There is step-by-step convergence that corresponds to the line of reasoning under level-1 thinking process.
In contrast to $L$, $H$ agents mention about fixed-strategy players, but as historical information becomes available, they no longer reason about their choices using the proportion of fixed-strategy players and focus more on the information about $\frac{2}{3}$ of the average.

\textit{Dynamic environment.}
%%In this subsection, LLM-based agents are playing against one another and they were not given any information on the proportion of LLM types in the group.
%%Since the algorithms are dynamically responsive, it would be harder to learn the proportion of agent types in period 2 even when past choices are revealed.
%%Therefore, in this case, there are more strategic uncertainty, and agents likely have to make use of other information to base their guesses on.
%%\subsubsection{Period 1}
%%\textbf{Understanding.}
In period 1, 
%%Once again, I verify agents' comprehension of game rules and objectives, which 
both $H$ and $L$ accurately recite game rules and objectives.
%%However, the main difference between the responses of different agent types is that the low types state straightaway that ``I think other players are most likely to choose numbers around 50, so I will choose 50" at this stage of eliciting understanding.
%~\palm~ responds with ``In this game, I need to choose a number between 0 and 100, and the winner is the person whose number is the closest to two thirds of the average of all chosen numbers. I think other players are most likely to choose numbers around 50, so I will choose 50." 
%\\ \ \\
%~\gptthree~ responds with ``The game involves choosing a number between 0 and 100, with the goal of being closest to two-thirds of the average of all chosen numbers. The winner receives a fixed prize, and in case of a tie, the prize is split among the winners. My primary goal is to maximize the possibility of winning the prize, and secondarily to maximize my own prize."
%\\ \ \\
%For all agents, when prompt to give an answer about what could be the popular number chosen, all of them answers $50$.
%%\\ \ \\
%%\textbf{Reasoning of Choice.}
Similar to the case with fixed-strategy opponents, when LLM-based agents are asked to state the reasoning of their choices, 
$L$ agents respond that they have chosen $50$ because they believe other players are most likely to choose numbers around $50$, which is the middle number and a safe bet.
%%thus choosing $50$ would offer them the best chance of winning.
%~\palm~ responds in period 1: ``I think other players are most likely to choose numbers around 50 because it is a middle number and it is a safe bet. If I choose a number that is too high or too low, I am less likely to win. By choosing 50, I am giving myself a good chance of winning."
%\\ \ \\
Meanwhile, $H$ agents also believe the popular answer is likely to be around $50$,
%%and many players choose it as a safe option,
but they respond to that by choosing $66.67$,
%%to maximize the likelihood of winning, which they stressed in the reasoning to be exactly two-thirds of the maximum possible value.
which again can be an indication that they are using the upper-bound as the focal point or they are following iterated elimination of dominated strategies.

In subsequent periods, as compared to the environment with fixed-strategy opponents, LLM-based agents in the game of LLMs vs. LLMs display slightly larger variability in the phrasing of their answers, while the content remain fairly consistent.
%%\\ \ \\
%%For instance, in period 2, a possible response from low type agents would be: ``In the last run, the average of all chosen numbers was $58.327999999999996$. The two thirds of the average is $38.88533333333333$. The winner was the player who chose $50$. I think other players are most likely to choose numbers around $50$ again in this run. To maximize my chance of winning, I will choose $66.67$, which is closer to the two thirds of the average than $50$. I also think that by choosing a number that is higher than $50$, I have a better chance of winning the prize if there is a tie."
When making their choices, $L$ agents take into consideration the average of all chosen numbers, $\frac{2}{3}$ of the average, as well as the winners' choice.
For $H$ agents,
%%a possible response would be: ``Based on the historical choices, the average of all chosen numbers in the previous run was $58.33$. To be closest to two-thirds of the average, I should aim for a number close to $38.89$. This is because two-thirds of $58.33$ is approximately $38.89$. By adjusting my choice to $38.89$, I increase my chances of being the closest to two-thirds of the average and winning the prize."
%%As illustrated, the high type agents 
they anchor their guesses to $\frac{2}{3}$ of the previous round average, complying with level-$1$ thinking.
They also appear to consider winner's strategy, where some of them indicate that that they are aligning their choices with the winning strategy from the previous round.
Based on overall line of reasoning, $L$ agents demonstrate mimicry of the winner's choice in previous round, and state that they believe the other players are most likely to choose the same winning number again.
%%This implies learning by imitation.
$H$ agents show adjustment in choices to $\frac{2}{3}$ of the previous period's average due to the following reasons:
(1) They incorporated information about winner's strategy, which is an indication of learning by imitation.
%%but they perceive the strategy to be selecting a number that is $\frac{2}{3}$ of the average instead of a strategy that is to select the winning number of the past round;
(2) By stating that they are aiming to be closer to $\frac{2}{3}$ of the past average implies adjustment according to level-$1$ reasoning, where the new guess is anchored to a new reference point.
(3) There is also a hint of outcome-based learning, where some mentioned they were not the closest to $\frac{2}{3}$ of average in the past round, and this propels a change in their strategy in the current round.
%\\
%Similar to the case with fixed-strategy opponents, there could be slight variations in terms of the choice of words and the decimal places in the guesses.
%In period 3, accounting for more decimal places and guessing $32.871$ instead of $32.87$ affect the agents adversely by making them further away from the average, whereas in period 4, guessing $23.418$ instead of $23.42$ is beneficial for the agents that account for longer string of 
(4) Lastly, a surprising thing that one agent mentioned was that ``considering the trend of decreasing choices in the previous runs, it seems reasonable to continue this trend and choose a lower number."
This highlights there could be learning based on pattern recognition.
\\ \ \\
Similar to the analysis before, since agents are allow to choose any number that falls within in the range, there is slight variations in the guesses chosen by $H$ agents due to token allocation.
%%(i.e. some chose $38.89$ and some $38.885$).
However, in this set-up, this distinction matters as the winner of the round that guesses with more or less decimal places could have won the game.
This property could have interesting implications.
It shows that even though information revealed are identical, there could be variations in information consideration and choices among homogeneous agents.
While such numerical variations are often trivial, having negligible impact on the determination of strategic levels, a small difference in choices could lead to a large difference in payoffs given the settings illustrated in this paper.
It is entirely possible that agents are deliberately choosing a number just slightly larger or smaller in order to beat the rest to be closer to two-thirds of the average,
or that agents are unconsciously selecting a longer string of decimal places, which could end up winning the game.
The settings with LLM-based agents most likely fall in the second category.
Since the information is feed to all agents at the beginning of each period, there are no distinction in what is being observed, therefore the difference lays in that some agents are able to process longer string of information, which technically boils down to token constraints.
However, this constraint can potentially relates to the processing capability of human subjects, illustrating the amount of ``attention" to the information given.
Having better ``attention" could imply incorporating longer string of information in decision-making, and thus higher payoffs in certain set-ups. 
%%Adopting this interpretation, there can be higher payoffs in certain set-ups even though the differences of number chosen between homogeneous agents are almost negligible.
Nonetheless, there can be instances where having more decimal place is detrimental to the outcome, such that the choice is further away from $\frac{2}{3}$ of the average than rounded-off numbers selected by players who do not pay as much ``attention".
Intriguingly, this potential opens up the study of attention in beauty contest game outcomes, which has yet to be addressed.

Return to Section~\ref{reasoningreturn}.

\end{document}